\documentclass[journal,final]{IEEEtran}
\usepackage{color,graphicx,amsmath}
\usepackage{eufrak}
\usepackage{enumerate}

\newtheorem{thm}{Theorem}
\newtheorem{cor}[thm]{Corollary}
\newtheorem{lem}[thm]{Lemma}
\newtheorem{prop}[thm]{Proposition}
\newtheorem{example}{Example}

\begin{document}
\title{Codes on Planar Graphs}
\author{\IEEEauthorblockN{Srimathy Srinivasan, Andrew Thangaraj\\}
\IEEEauthorblockA{Department of Electrical  Engineering\\
Indian Institute of Technology, Madras\\
Email: andrew@iitm.ac.in}}

\maketitle

\begin{abstract}
Codes defined on graphs and their properties have been subjects of intense recent research. On the practical side, constructions for capacity-approaching codes are graphical. On the theoretical side, codes on graphs provide several intriguing problems in the intersection of coding theory and graph theory. In this paper, we study codes defined by planar Tanner graphs. We derive an upper bound on minimum distance $d$ of such codes as a function of the code rate $R$ for $R \ge 5/8$. The bound is given by 
$$d\le \left\lceil \frac{7-8R}{2(2R-1)} \right\rceil  + 3\le 7.$$
Among the interesting conclusions of this result are the following: (1) planar graphs do not support asymptotically good codes, and (2) finite-length, high-rate codes on graphs with high minimum distance will necessarily be non-planar. 
\end{abstract}

\section{Introduction}
The spectacular success of codes on graphs has resulted in immense recent research activity on the practical and theoretical aspects of graphical codes. On the practical side, the powerful notion of representing parity constraints on Tanner graphs \cite{Gallager:1963xy}\cite{Tanner:1981gd} has resulted in tremendous simplifications in the construction and implementation of capacity-approaching codes for various channels. On the theoretical side, the interplay of graph theory and coding theory has resulted in many intriguing problems. 

In this paper, we are concerned with codes that are defined by planar Tanner graphs. Specifically, we study the minimum distance of codes that have a planar Tanner graph. Planarity of a graph, a classic notion in graph theory, allows for the embedding or rendering of a graph as a picture on a two-dimensional plane with no two edges intersecting. Specific examples of such graphs are trees and graphs with non-overlapping cycles. Interestingly, both these types of graphs have been shown to correspond to codes with poor minimum distance properties \cite{Etzion:1999jx}\cite{Srimathy:2008qy}. In this paper, we show similar properties for high-rate codes that have planar Tanner graphs.

Specifically, the main result of this paper is that a code of rate $R\ge5/8$ with a planar Tanner graph has minimum distance bounded as 
$$d\le \left\lceil \frac{7-8R}{2(2R-1)} \right\rceil  + 3\le 7,$$
where $\lceil x\rceil$ (for a real number $x$) is the smallest integer greater than or equal to $x$. Note that the result holds for any blocklength. Hence, non-planarity is vital for large minimum distance at high rates. This result provides justification for many known results on codes on highly non-planar graphs with large minimum distance \cite{Tanner:2001di}, and suggests methods for other possible constructions. 

The method of proof is novel and involves several steps. A given planar Tanner graph is modified through a series of construction steps to a planar Tanner graph with maximum bit node degree 3. For the modified graph, the existence of low-weight codewords is shown by an averaging argument. The existence is then extended to the original Tanner graph. 

The rest of the paper is organized as follows. The construction of the modified Tanner graph is presented in Section \ref{sec:constructions}. A simple version of the main result is proved in Section \ref{sec:distance-rate-bounds} for the sake of clarity in exposition. A complete proof of the bound on minimum distance for codes on planar graphs is given in Section \ref{sec:proof-main-result}. Finally, concluding remarks are made in Section \ref{sec:conclusion}.
\section{Constructions}
\label{sec:constructions}
Consider a Tanner graph $G$ defining a linear code. The vertex set of $G$ is denoted $V(G)=V^b\cup V^c$, where $V^b$ and $V^c$ denote the set of bit and check nodes of the $G$. We will assume that the rate of the code defined by $G$ is $R=1-|V^c|/|V^b|$. Let $V^b_i$ denotes the set of degree-$i$ bit nodes. The edge set of $G$ is denoted $E(G)\subseteq V^b\times V^c$, and an edge of $G$ connecting bit node $v_b$ to check node $v_c$ is denoted $(v_b,v_c)$. 

For a set of bit nodes $V^b_*\subseteq V^b$, $\mathcal{N}(V^b_*)$ denotes the set of check nodes connected to $V^b_*$ i.e. $\mathcal{N}(V^b_*)=\cup_{v_b\in V^b_*}\{v_c:(v_b,v_c)\in E(G)\}$. The degree of a bit node $v_b$ is $|\mathcal{N}(v_b)|$. For $V^c_*\subseteq V^c$, the set of induced bit nodes $\mathcal{I}(V^c_*)$ denotes the bit nodes whose neighbors are subsets of $V^c_*$ i.e. $\mathcal{I}(V^c_*)=\{v_b:\mathcal{N}(v_b)\subseteq V^c_*\}$. The following proposition (stated without proof) connects subsets of check nodes and their induced bit nodes to the minimum distance of the code.
\begin{prop} 
\label{prop:dmin}
Consider $V^c_*\subseteq V^c$ in a Tanner graph $G$ defining a code with minimum distance $d$. If $\mathcal{I}(V^c_*)> V^c_*$, then  $d\le|V^{c}_*| +1$.
\end{prop}

Following Proposition \ref{prop:dmin}, a subset of check nodes $V^c_*$ is said to be {\it codeword-supporting} whenever $\mathcal{I}(V^c_*)> V^c_*$.

In this paper, we provide bounds for the minimum distance of Tanner graphs that are planar i.e. Tanner graphs that can be embedded in a plane with no two edges intersecting \cite{Bondy:1976hl}. All Tanner graphs in the rest of the paper will be planar with a fixed embedding. For planar Tanner graphs, we use Proposition \ref{prop:dmin} for bounding minimum distance by showing the existence of suitable codeword-supporting subsets of check nodes. For this purpose, we define a new planar graph involving the check nodes of the given planar Tanner graph.
 
\subsection{Check graph of a planar Tanner graph}
\begin{figure}[ht!]
\centering
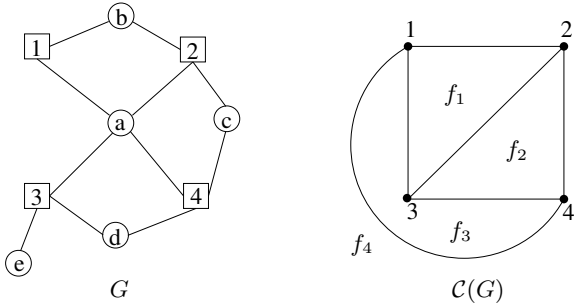
\caption{Construction of check graph I.}
\label{fig:cg}
\end{figure}

Given a planar Tanner graph $G$, the \textit{check graph} of $G$, denoted $\mathcal{C}(G)$, is a planar graph with vertex set $V^c$ (the set of check nodes of $G$). We use an embedding of $G$ in a plane, and place the nodes of $\mathcal{C}(G)$ in an isomorphic plane at the same locations as the check nodes of $G$ in the original plane. To aid in the construction, we identify the locations of the bit nodes of $G$ in the plane of $\mathcal{C}(G)$. The edges of $\mathcal{C}(G)$ are constructed as follows:
\begin{enumerate}
\item Consider a bit node $v_b\in V^b$ with degree $\lambda>1$ and $\mathcal{N}(v_b)=\{v_{c,0},v_{c,1},\cdots,v_{c,\lambda-1}\}$ labelled in a clockwise sequence in the planar embedding i.e. no edge out of $v_b$ lies in \angle$v_{c,i}v_bv_{c,(i+1)_\lambda}$ ($(x)_{\lambda}$ denotes $x\mod \lambda$). Add edges $(v_{c,i},v_{c,(i+1)_\lambda})$ for $0\leq i\leq \lambda-1$ to form a simple cycle enclosing the location of $v_b$ in the plane of $\mathcal{C}(G)$. 
\item In Step 1, edges causing a face enclosed by two edges should not be added.
\item Add more edges to make the graph maximal planar (a planar graph is maximal if one more edge will make the graph non-planar \cite{Bondy:1976hl}).
\end{enumerate}
The construction of the check graph is illustrated in Fig. \ref{fig:cg}. 
In Fig. \ref{fig:cg}, the nodes of the check graph $\mathcal{C}(G)$ are the check nodes of $G$ denoted $\{1,2,3,4\}$. In Step 1, for bit node `a' of $G$, we connect the nodes $\{1,2,4,3\}$ of $\mathcal{C}(G)$ in a cycle. Step 1, for the nodes `b', `c' and `d' of $G$ with degree larger than 1, results in faces with two edges. Hence, no other edges are added to $\mathcal{C}(G)$ as per Step 2. In Step 3, the edges (2,3) and (1,4) are added to make the check graph maximal planar. Note that there are four faces in $\mathcal{C}(G)$, labelled $f_1$, $f_2$, $f_3$ and $f_4$ in Fig. \ref{fig:cg}. The face $f_4$ is the exterior or external face.

In general, maximal planarization in Step 3 is not unique. Hence, there can be many check graphs corresponding to a single Tanner graph. We fix one such check graph and call it the check graph of $G$. 

The following correspondences between a planar Tanner graph $G$ and its check graph $\mathcal{C}(G)$ are vital for the minimum distance bounds.
\begin{itemize}
\item In Step 1, a degree-3 bit node $v_b$ of $G$ maps to a triangular face in $\mathcal{C}(G)$ connecting the three check nodes in $\mathcal{N}(v_b)$. Hence, we say that a degree-3 bit node is ``identified" with a face in $\mathcal{C}(G)$. In some cases, this can be the external face.
\item In Step 1, a degree-2 bit node $v_b$ of $G$ maps to an edge in $\mathcal{C}(G)$ connecting the two check nodes in $\mathcal{N}(v_b)$. We say that a degree-2 bit node is ``identified" with an edge in $\mathcal{C}(G)$.
\item A degree-1 bit node $v_b$ of $G$ does not result in any edges, but $v_b$ is represented by the one check node $\mathcal{N}(v_b)$ in $\mathcal{C}(G)$. We say that a degree-1 bit node $v_b$ is ``identified" with the check node $\mathcal{N}(v_b)$. 
\item In Step 1, a bit node of degree $\lambda>3$ results in a face enclosed with $\lambda$ edges. In Step 3, maximum planarization converts such a face into $\lambda-2$ triangular faces.
\end{itemize}
In the example of Fig. \ref{fig:cg}, the degree-4 bit node of $G$ results in two triangular faces $f_1$ and $f_2$ in $\mathcal{C}(G)$. The check node 3 corresponds to the degree-1 bit node, while the edges $\{(1,2), (2,4), (3,4)\}$ correspond to the degree-2 bit nodes.

\begin{figure}[ht]
  \centering
  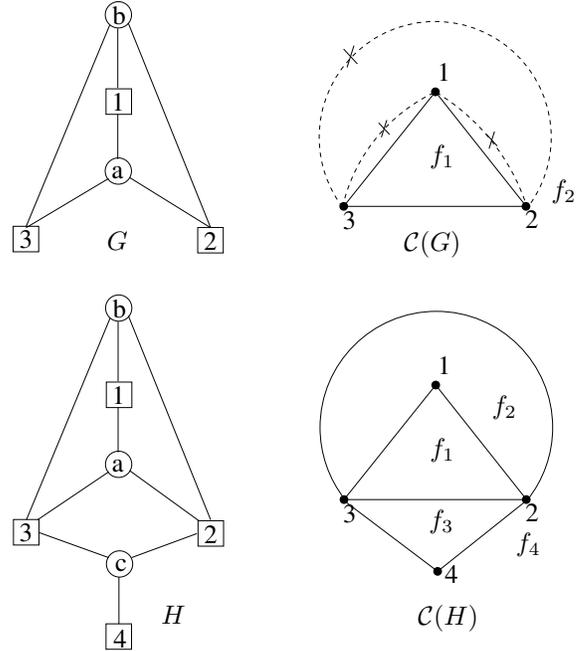
  \caption{Construction of check graph II.}
  \label{fig:ex2}
\end{figure}

Two more examples to illustrate the construction of the check graph are shown in Fig. \ref{fig:ex2}. 
For the graph $G$ in Fig. \ref{fig:ex2}, we get a triangle around the location of bit node `a' in Step 1. For bit node `b', the dotted lines show the possible edges in Step 1. However, no new edges are added as they result in faces enclosed by two edges. Notice that the circular dotted line would have resulted in the external face being ``enclosed'' by two edges. In $G$, the internal triangular face $f_1$ corresponds to the degree-3 bit node `a', while the external triangular face $f_2$ corresponds to `b'.

For the graph $H$ in Fig. \ref{fig:ex2}, two triangles are added around the bit nodes `a' and `c' in Step 1. Note that the edge from check node 3 to check node 2 for bit node `b' needs to be drawn in a circular fashion to enclose the location corresponding to `b'.

Since $\mathcal{C}(G)$ is maximal planar, by standard results in graph theory \cite{Bondy:1976hl}, we know that there are $2|V^c|-4$ faces in $\mathcal{C}(G)$ and all faces are triangular (enclosed by three edges). Also, since faces in $\mathcal{C}(G)$ result from bit nodes of degree at least 3, we see that the maximum number of bit nodes of degree 3 in a planar Tanner graph is limited to $2|V^c|-4$. 

The faces of $\mathcal{C}(G)$ are denoted $F(\mathcal{C}(G))$. A face $f\in F(\mathcal{C}(G))$ is enclosed by three edges connecting three check nodes of $G$. The three check nodes of $G$ that form $f$ are denoted $V^c(f)$. In the example of Fig. \ref{fig:cg}, we have $V^c(f_1)=\{1,2,3\}$, $V^c(f_2)=\{2,3,4\}$, $V^c(f_3)=\{1,3,4\}$, and $V^c(f_4)=\{1,2,4\}$.

\subsection{Prelude}
\label{sec:prelude}
To illustrate the usefulness of the check graph, we now present a bound on minimum distance of rate $\ge7/8$ codes with a planar Tanner graph whose maximum bit node degree is 3. 
\begin{prop}
Let $G$ be a planar Tanner graph with $n$ bit nodes and $m$ check nodes defining a code with rate $R=1-m/n\ge7/8$ and minimum distance $d$. Let the maximum degree of a bit node in $G$ be 3. Then, $d\le3$. 
\label{prop:prelude}
\end{prop}
\begin{IEEEproof}
Let $f_i$, $1\le i\le2m-4$, be the faces of the check graph $\mathcal{C}(G)$. Let $w_i=|\mathcal{I}(V^c(f_i))|$ be the number of bit nodes induced by the set of check nodes $V^c(f_i)$ forming the face $f_i$. Let $f_k$ be the face such that $w_k\ge w_i$, $1\le i\le2m-4$. 

Since each bit node of $G$ has degree at most 3, it is induced at least once by some face $f_i$ in $\mathcal{C}(G)$. So, we have
$$w_1 + w_2 + \cdots+ w_{2m-4} \ge n.$$
Since $w_k\ge w_i$ and $n\ge 8m$, we simplify as follows.
\begin{align*}
w_k(2m-4) & \ge n,\\
w_k \ge \frac{n}{2m-4} & > \frac{n}{2m} \ge 4.
\end{align*}
Hence, $w_k = |\mathcal{I}(V^c(f_k))| \ge 5$ and $|V^c(f_k)| = 3$. Thus, there exists a $(\ge 5, 3)$ subcode of the original code. This implies that $d\le3$.
\end{IEEEproof}

The above proof uses the faces of $\mathcal{C}(G)$ to construct codeword-supporting set of check nodes in the original Tanner graph $G$. An averaging argument is used to show the existence of the codeword-supporting set. These themes will be used and extended in the remainder of this paper to prove more bounds on the minimum distance of codes with a planar Tanner graph. 

Another crucial assumption in Proposition \ref{prop:prelude} is on the maximum bit node degree. This assumption will be relaxed through another construction called the {\it check inverse}.

\subsection{Check inverse of a planar Tanner graph}
\begin{figure}[ht!]
\centering
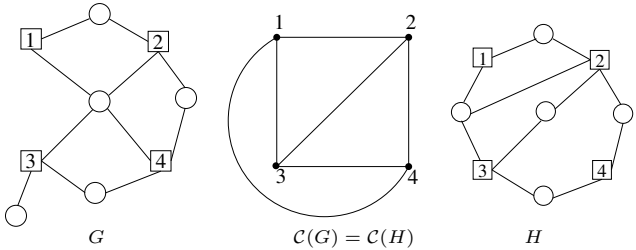
\caption{One check graph for multiple Tanner graphs.}
\label{fig:mcg}
\end{figure}

The same check graph can result from several planar Tanner graphs. Fig. \ref{fig:mcg} illustrates one such example where two Tanner graphs $G$ and $H$ result in the same check graph $\mathcal{C}(G)$.

Given a check graph $\mathcal{C}(G)$ of a planar Tanner graph $G$, we can construct a special planar Tanner graph $G'$ with maximum bit node degree 3 such that $\mathcal{C}(G')=\mathcal{C}(G)$. The construction of this special Tanner graph, which we call the {\it check inverse} of $G$, is described next.

Given a planar Tanner graph $G$, the check inverse of $G$, denoted $G'$, is a planar Tanner graph with check node set $V'^{c} = V^{c}$ and bit node set $V'^{b}=V'^{b}_3\cup V'^{b}_2\cup V'^{b}_1$ constructed as follows:
\begin{enumerate}
\item $V'^{b}_3\cong F(\mathcal{C}(G))$. A bit node $v_b\in V'^{b}_3$ corresponds to a face $f\in F(\mathcal{C}(G))$, and is connected to the three check nodes in $V^c(f)$ that form the face $f$ in $\mathcal{C}(G)$. Bit nodes in $V'^{b}_3$ have degree 3, and $|V'^{b}_3|=2|V^c|-4$.
\item The set $V_2^{'b} \bigcup V_1^{'b}$ is an arbitrary subset of $V_2^b \bigcup V_1^b$ (the set of degree-2 and degree-1 bit nodes of $G$) of size 
$$|V'^{b}_2 \bigcup V_1^{'b}|=[|V^b|-(2|V^c|-4)]^+$$ 
where $[x]^+ = \max(x,0)$. A bit node $v_b \in V_2^{'b} \bigcup V_1^{'b} \subseteq V^b_2 \bigcup V_1^b$ is connected to the check node(s) in $\mathcal{N}(v_b)$.
\end{enumerate}

\begin{example}
\label{ex:GtoG'}
The construction of the check inverse for a planar Tanner graph $G$ is illustrated in Fig. \ref{fig:chkinv}. 
\begin{figure}[ht!]
\centering
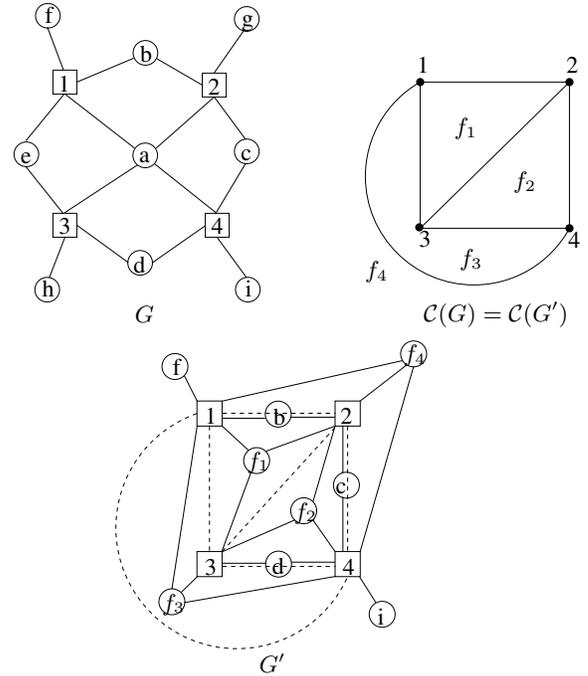
\caption{Construction of check inverse.}
\label{fig:chkinv}
\end{figure}
The Tanner graph $G$ has 9 bit nodes (1 of degree 4, 4 of degree 2, 4 of degree 1) and 4 check nodes. The check graph $\mathcal{C}(G)$ has 4 nodes, corresponding to the check nodes of $G$, and 4 faces (3 interior and 1 exterior) enclosed by three edges each. The check inverse $G'$ has 4 check nodes corresponding to the check nodes of $G$ or the nodes of $\mathcal{C}(G)$. The 4 bit nodes of degree 3 in $G'$ correspond to the 4 faces of $\mathcal{C}(G)$ from Step 1 of the construction of $G'$. These nodes are labelled with the labels of the faces in $\mathcal{C}(G)$. In Step 2, 5 bit nodes of degree-1 and 2 (the nodes `b', `c', `d', `f', `i') are added to $G'$ with connections according to the corresponding connections in $G$. In Fig. \ref{fig:chkinv}, the dotted lines in $G'$ are the edges of $\mathcal{C}(G)=\mathcal{C}(G')$.
\end{example}

The following properties of the check inverse of a planar Tanner graph are important for future constructions:
\begin{itemize}
\item If $|V^b|>2|V^c|-4$, the check inverse $G'$ has the same number of check nodes and bit nodes as $G$. If the rate of the code defined by $G$ is greater than 1/2, we have $|V^b|>2|V^c|-4$.
\item For rate greater than 1/2, we have 
  \begin{equation}
    \label{eq:deg12}
|V'^b_1|+|V'^b_2|=|V^b|-(2|V^c|-4).    
  \end{equation}
  From now on, we will restrict ourselves to planar Tanner graphs $G$ that define codes of rate greater than 1/2 so that (\ref{eq:deg12}) always holds.
\item The check graph of $G'$ is same as $\mathcal{C}(G)$ i.e, $\mathcal{C}(G')=\mathcal{C}(G)$. But each face in $\mathcal{C}(G')$ corresponds to a degree 3 bit node in $G'$ unlike in $\mathcal{C}(G)$ and $G$. 
\end{itemize}

The check inverse plays a crucial role in the minimum distance bounds. Distance bounds will first be shown for the code represented by the Tanner graph $G'$. Then, the same bound will be seen to hold for the original graph $G$.

\subsection{Dual of check graph and codeword-supporting subgraphs}
\label{sec:dual}
Since the check graph $\mathcal{C}(G)$ is planar, we can define its dual as defined for any planar graph \cite{Bondy:1976hl}. Since we are working on a particular embedding, the dual graph is unique. 

The dual of $\mathcal{C}(G)$, denoted as $\mathcal{C}^\dagger(G)$, is a planar graph with vertex set $V^\dagger(G)$ that has a one-to-one correspondence with $F(\mathcal{C}(G))$. Two vertices of $\mathcal{C}^\dagger(G)$ are joined by an edge whenever the corresponding faces of $\mathcal{C}(G)$ share an edge. Since $\mathcal{C}(G)$ is maximal planar with $|V^c|$ vertices and $2|V^c|-4$ triangular faces, $\mathcal{C}^\dagger(G)$ has $2|V^c|-4$ vertices each of degree 3. Since there is a one-to-one correspondence between edges in a planar graph and its dual, let us denote the edge corresponding to $e$ in $\mathcal{C}(G)$ as $e^\dagger$ in $\mathcal{C}^\dagger(G)$.  

If there are multiple edges between any two nodes of $\mathcal{C}^\dagger(G)$, we can show that the minimum distance of the code represented by $G$ is at most 2 (See Section \ref{sec:girth-condition} for a proof). So, we consider planar Tanner graphs $G$ and check graphs $\mathcal{C}(G)$ such that there are no multiple edges in $\mathcal{C}^\dagger(G)$. 

Subgraphs of the dual of the check graph play an important role in determining the existence of low-weight codewords in the code (or small codeword-supporting subsets of check nodes in the code's Tanner graph). The basic idea is the following. Using a vertex-induced subgraph of the dual of the check graph $\mathcal{C}^\dagger(G)$, we define subsets of check nodes of $G$ and study when they are codeword-supporting. Let $U\subseteq V^\dagger(G)\cong F(\mathcal{C}(G))$ induce a subgraph $\mathcal{C}_U^\dagger(G)$ of $\mathcal{C}^\dagger(G)$. Because of the congruence, we will denote a vertex of $\mathcal{C}^\dagger(G)$ as a face $f\in F(\mathcal{C}(G))$. Hence, $U\subseteq F(\mathcal{C}(G))$. For each subset $U$, we associate a subset of check nodes $V^c_U$ of $G$ as given below:
\begin{equation}
  \label{eq:vcu}
V^c_U=\cup_{f\in U}V^c(f),  
\end{equation}
where $V^c(f)$ (as before) is the set of three nodes that form the face $f$ in $\mathcal{C}(G)$. Hence, every vertex subset $U$ in $\mathcal{C}^\dagger(G)$ corresponds to a subset of check nodes $V^c_U$ in $G$. We will prove existence of codeword-supporting subsets among the sets $V^c_U$ produced by different $U$.

The subgraph of  $\mathcal{C}(G)$ induced by $V^c_U$ is denoted $\mathcal{C}_U(G)$. In spite of the notation, note that $\mathcal{C}_U^\dagger(G)$ is not necessarily the planar dual of $\mathcal{C}_U(G)$.

As before, the set of bit nodes induced by $V^c_U$ in the Tanner graph $G$ is denoted $\mathcal{I}(V^c_U)$. The {\it weight} of an induced subgraph $\mathcal{C}_U^\dagger(G)$, denoted $wt(\mathcal{C}_U^\dagger(G))$ or $wt(U)$, is defined as $wt(U) = |\mathcal{I}(V^c_U)|$, the number of induced bit nodes of $V^c_U$. An induced subgraph $\mathcal{C}_U^\dagger(G)$ of $\mathcal{C}^\dagger(G)$ is said to be {\it codeword-supporting} if $wt(U) > |V^c_U|$. 

\begin{example}
Consider a planar Tanner graph $G$ with $V^b = \{\text{a,b,c,d,e,f,g}\}$ and $V^c = \{1,2,3,4,5\}$ as shown in Fig. \ref{fig:induced_bit}. 
\begin{figure}[ht!]
\centering
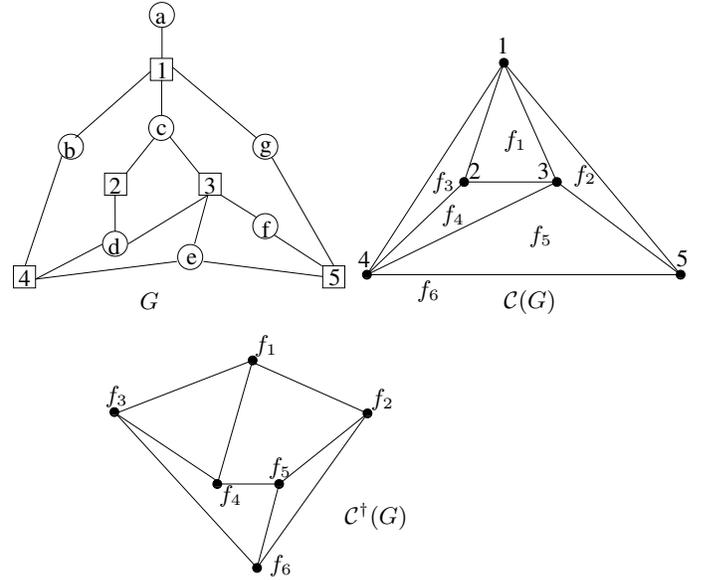
\caption{Dual of the check graph.}
\label{fig:induced_bit}
\end{figure}
The check graph $\mathcal{C}(G)$ has 5 vertices corresponding to 5 check nodes of $G$ and 6 faces (including the external face). The 6 vertices in  dual $\mathcal{C}^\dagger(G)$ are labelled according to the corresponding faces in  $\mathcal{C}(G)$. Let $U=\{f_1,f_4\} \subset V^\dagger(G)$. Then, $V^c_U = \{1,2,3,4\}$ and  $\mathcal{C}_U(G) = f_1 \bigcup f_3 \bigcup f_4$. The bit nodes induced by  $V^c_U$, $\mathcal{I}(V^c_U) = \{\text{a,b,c,d}\}$ and hence $wt(U) = 4$.

In Fig. \ref{fig:induced_bit}, the degree-3 bit nodes of $G$ $\{\text{c,d,e}\}$ are identified with the respective faces $\{f_1,f_4,f_5\}$ in the check graph $\mathcal{C}(G)$ and the corresponding vertices in the dual of the check graph $\mathcal{C}^\dagger(G)$. The degree-2 bit nodes of $G$ $\{\text{b,f,g}\}$ are identified with the respective edges $\{(1,4), (3,5), (1,5)\}$ in $\mathcal{C}(G)$ and the corresponding edges $\{(f_3,f_6), (f_2,f_5), (f_2,f_6)\}$ in $\mathcal{C}^\dagger(G)$. The degree-1 bit node $\{\text{a}\}$ of $G$ is identified with node 1 of $\mathcal{C}(G)$ and the external face of $\mathcal{C}^\dagger(G)$.
\label{ex:dual-check-graph}
\end{example}

In the main result of this paper, we establish the existence of codeword-supporting subgraphs induced by a small subset $U$ in the dual of the check graph of a planar Tanner graph. The next two propositions show that the size of $V^c_U$ and the minimum distance of the code are bounded by the size of $U$ that induces a codeword-supporting subgraph in $\mathcal{C}^\dagger(G)$. Hence, small codeword-supporting subgraphs in the dual of the check graph result in low-weight codewords in the code.
\begin{prop}Let $\mathcal{C}_U^\dagger(G)$ be a connected induced subgraph of $\mathcal{C}^\dagger(G)$ induced by a proper subset $U\subset F(\mathcal{C}(G))$. Then, for the subset of vertices $U$,
$$|V^c_U| \le |U| + 2 - c(U),$$
where $c(U)$ is the number of simple cycles in $\mathcal{C}_U^\dagger(G)$.
\label{prop:vcu}
\end{prop}
\begin{IEEEproof} 
As shown in Appendix \ref{sec:numbering}, the subgraph $\mathcal{C}_U^\dagger(G)$ can be constructed by adding nodes one at a time from the set $U$ in a suitable order. The order is such that, at each step of the construction, the most recently added node has degree 1 or 2 after inclusion. Hence, the resulting subgraph after each step is connected. With this particular ordering of nodes of $U$, we will prove the proposition by induction.

Let $|U| = 1$. Then, $|V^c_U| = 3 = |U| + 2 - 0$. Hence, the proposition is true for $|U| =1$. Assume that it is true for the subgraph induced by $P$ nodes where $P \subset U$ i.e. 
$$
|V^c_P|  \le |P| +2 - c(P),
$$
where $c(P)$ is the number of simple cycles in $\mathcal{C}_P^\dagger(G)$. We will prove that the result holds when a new node $w$ is added. Consider the subgraph $\mathcal{C}_W^\dagger(G)$ induced by $W = P \bigcup w$. By the ordering of the nodes, $w$ has degree 1 or 2 in $\mathcal{C}_W^\dagger(G)$.

\noindent\textit{Case} 1: $w$ has degree 1.

Since $w$ is connected to $\mathcal{C}_P^\dagger(G)$ by exactly one edge, the face in $\mathcal{C}_W(G)$ corresponding to $w$ results in the addition at most one new node to $\mathcal{C}_P(G)$. Also, the addition of $w$ does not create a new cycle in $\mathcal{C}_W^\dagger(G)$. Hence, $|V^c_W| \le |V^c_P| + 1 \le |P| +2 - c(P) + 1 = |W| +2 - c(W)$, where $c(W)$ is the number of simple cycles in $\mathcal{C}_W^\dagger(G)$ which is equal to $c(P)$. 

\noindent\textit{Case} 2: $w$ has degree 2

Since $w$ is connected to $\mathcal{C}_P^\dagger(G)$ by two edges, the nodes at the boundary of the face corresponding to $w$ in $\mathcal{C}_W(G)$ are already present in $\mathcal{C}_P(G)$. Thus number of nodes in $\mathcal{C}_P(G)$ is same as that in $\mathcal{C}_W(G)$. Also, the addition of $w$ increases the number of cycles in the resulting subgraph by 1. This is because $w$ connects two nodes in $P$ that are already connected in $\mathcal{C}_P^\dagger(G)$. Hence, $|V^c_W| = |V^c_P|  \le |P|  +2 - c(V) + 1 -1 = |W| +2 - c(W)$ where $c(W)$ is the number of simple cycles in $\mathcal{C}_W^\dagger(G)$.

The two cases are illustrated in Fig. \ref{fig:check_count}. 
\begin{figure}[ht!]
\centering
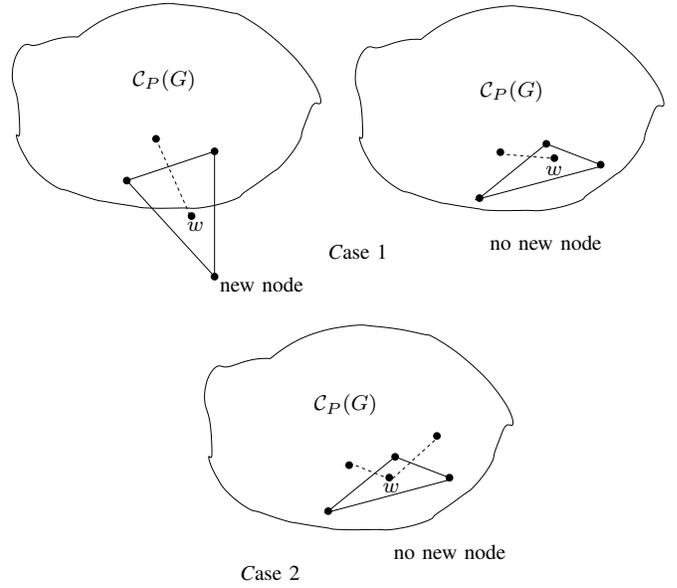
\caption{One step in inductive construction of $\mathcal{C}_U^\dagger(G)$.}
\label{fig:check_count}
\end{figure}
The dotted lines in Fig. \ref{fig:check_count} are the edges induced in $\mathcal{C}^\dagger(G)$ by the addition of the node $w$.

Thus by induction, 
$$|V^c_U| \le |U| + 2 - c(U).$$
\end{IEEEproof}

\begin{prop}Let $G$ be a planar Tanner graph of a code with minimum distance $d$. If there is a codeword-supporting connected subgraph of $\mathcal{C}^\dagger(G)$ induced by a proper subset $U$, then
$$d\le |U|+3.$$
\label{prop:cu}
\end{prop}
\begin{IEEEproof}
Since $\mathcal{C}_U^\dagger(G)$ is codeword-supporting, $wt(U) = |\mathcal{I}(V^c_U)|> |V^c_U|$. By Proposition \ref{prop:dmin}, $d \le |V^c_U| +1$. Since $\mathcal{C}_U^\dagger(G)$ is connected, by Proposition \ref{prop:vcu}, $|V^c_U| = |U| + 2 -c\le |U| + 2$. Therefore, $d \le |U|+3$.
\end{IEEEproof}
Therefore, to bound minimum distance, we search for subgraphs of $\mathcal{C}^\dagger(G)$ on minimum number of vertices that are codeword-supporting. 

\section{Distance-Rate Bounds}
\label{sec:distance-rate-bounds}
The main result of this paper is the following theorem.
\begin{thm}
Let $G$ be a planar Tanner graph representing a code with rate $R\ge5/8$ and minimum distance $d$. Then,
$$d\le \left\lceil \frac{7-8R}{2(2R-1)} \right\rceil  + 3.$$
\label{thm:main}
\end{thm}
The proof involves multiple steps. In the first step, we construct the check inverse $G'$, check graph $\mathcal{C}(G')=\mathcal{C}(G)$ and its dual $\mathcal{C}^\dagger(G')$ as discussed in Section \ref{sec:constructions}. In the second step, the distance bound of Theorem \ref{thm:main} is shown for the code corresponding to the check inverse $G'$ by proving the existence of a suitable codeword-supporting subgraph in $\mathcal{C}^\dagger(G')$. In the third and final step, the same bound is shown to hold for $G$ by a series of graph manipulations.

For clarity of explanation, we first show the second and third steps in the proof for a weaker version of Theorem \ref{thm:main}. For the weaker version, the codeword-supporting subgraph of $\mathcal{C}^\dagger(G')$ is simply an edge. However, the important ideas in the general proof are present in the weaker version as well. A general proof of Theorem \ref{thm:main} is presented later in Section \ref{sec:proof-main-result}.

\subsection{Illustrative proof}
A weaker version of Theorem \ref{thm:main} is the following.
\begin{thm} Let $G$ be a planar Tanner graph representing a code with rate $R\ge11/16$ and minimum distance $d$. Then, $d\le5$. 
\label{thm:weak}
\end{thm}

We first prove a few lemmas that are used in the final proof. Using the constructions in Section \ref{sec:constructions}, let $G'$ be the check inverse of $G$. Let $\mathcal{C}(G')$ be the check graph of $G'$ and let its dual be $\mathcal{C}^\dagger(G')$. 

\subsubsection{Codeword-supporting subgraph for $G'$}
\begin{lem}  Let $G'$ be a check inverse of $G$ supporting a code of rate $R \ge 11/16$. Then, there is an edge in $\mathcal{C}^\dagger(G')$ that is codeword-supporting.
\label{lemma:1}
\end{lem}
\begin{IEEEproof} Consider an edge $e^\dagger=(f_1,f_2)\in E^\dagger(G')$, where $E^\dagger(G')$ denotes the edge set of $\mathcal{C}^\dagger(G')$. Since $\mathcal{C}(G')$ is maximal planar on $|V^c|$ vertices, we have $|E^\dagger(G')|=3|V^c|-6$. We set $U=\{f_1,f_2\}$ and get $\mathcal{C}_U^\dagger(G')=e^\dagger$ as shown in Fig \ref{fig:edge_cs}. For simplicity, the set $U$ is replaced with $e^\dagger$ in the notation. For instance, $V^c_U$ will be denoted $V^c_{e^\dagger}$ and so on.

\begin{figure}[ht!]
\centering
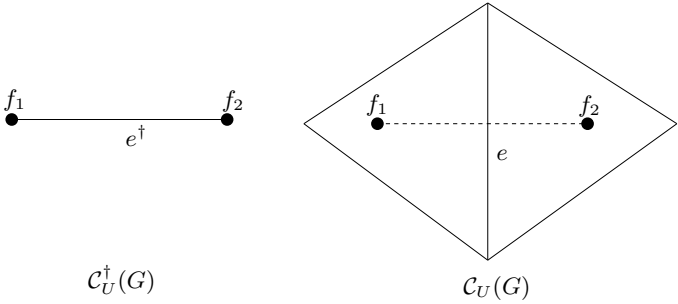
\caption{Edge as the induced subgraph.}
\label{fig:edge_cs}
\end{figure}

Consider the following summation:
\begin{equation}
  \label{eq:y_p2}
\mathcal{Y}(G')= \sum_{e^\dagger \in E^\dagger(G')} \left(wt(e^\dagger) - |V^c_{e^\dagger}|\right).
\end{equation}
We will show that $\mathcal{Y}(G')>0$, which implies that there exists an edge $e^\dagger$ such that $wt(e^\dagger) > |V^c_{e^\dagger}|$ proving the lemma. 

From Fig. \ref{fig:edge_cs}, $|V^c_{e^\dagger}|=4$ for all $e^\dagger$. So, to evaluate $\mathcal{Y}(G')$, we write $\sum_{e^\dagger \in E^\dagger(G')} wt(e^\dagger)$ as follows:
$$\sum_{e^\dagger \in E^\dagger(G')} wt(e^\dagger)=\sum_{v_b\in V'^b}q(v_b),$$
where $q(v_b)$ is the number of times a node $v_b$ of $G'$ is induced by $V^c_{e^\dagger}$, $e^\dagger \in E^\dagger(G')$. Let
\begin{align}
\eta_i = \sum_{{v_b \in V_i^{'b}}} q(v_b) \label{eta}
\end{align}
where $V_i^{'b}$ is the set of degree-$i$ bit nodes in $G'$. Then,
\begin{equation*}
   \sum_{e^\dagger \in E^\dagger(G')} wt(e^\dagger)=\eta_1 + \eta_2 + \eta_3
\end{equation*}

We can now evaluate $q(v_b)$ in the terms $\eta_i$ for $i=1,2,3$. A degree-3 bit node $v_b\in V'^b_3$ corresponds to a triangular face in $\mathcal{C}(G')$, which corresponds to a degree-3 node  $f \in\mathcal{C}^\dagger(G')$. Whenever an edge $e^\dagger$ is incident on $f$, the node $v_b$ will be induced by $V^c_{e^\dagger}$. Since there are 3 edges incident on any node in $\mathcal{C}^\dagger(G')$, $q(v_b) = 3$ for $v_b\in V'^b_3$.

A degree-two bit node $v_b\in V'^b_2$ is identified with an edge in $\mathcal{C}(G')$. Note that this edge is common to two faces, say $f_1$ and $f_2$ in $\mathcal{C}(G')$. Let $f_1$ and $f_2$ be the corresponding nodes in $\mathcal{C}^\dagger(G')$. Then, $v_b$ is induced by $V^c_{e^\dagger}$, whenever $e^\dagger$ is incident to $f_1$ or $f_2$. Since there are 5 edges incident on two neighboring nodes $f_1$ and $f_2$ in $\mathcal{C}^\dagger(G')$, $q(v_b) = 5$ for $v_b\in V'^b_2$. This is illustrated in Fig \ref{fig:q_deg2_p2}.

\begin{figure}[ht!]
\centering
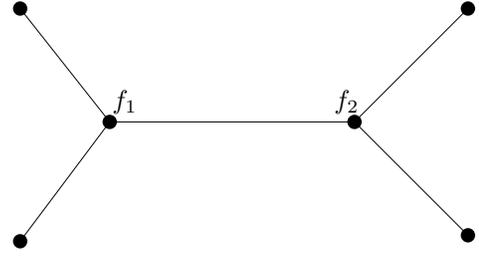
\caption{Computing $q(v_b)$ for a degree 2 bit node $v_b$.}
\label{fig:q_deg2_p2}
\end{figure}

A degree-1 bit node $v_b\in V'^b_2$ in $G'$ is identified by a check node to which it is connected to in $G'$. This check node corresponds to a face in $\mathcal{C}^\dagger(G')$. The node $v_b$ is induced by $V^c_{e^\dagger}$, whenever $e^\dagger$ is incident on one or more vertices of the face. Since there are at least 3 vertices in a face, note that $q(v_b)$ for a degree-1 bit node $v_b$ is greater than $q(v'_b)$ for  a degree-2 bit node $v'_b$. Thus,
\begin{eqnarray*}
\sum_{e^\dagger \in E^\dagger(G')} wt(e^\dagger) &\ge& 3|V_3^{'b}|  + 5 |V_2^{'b}| + 5|V_1^{'b}|,\\
&=& 3(2|V'^c|-4) + 5(|V'^b| - (2|V'^c|-4)),\\
&=&5|V'^b|-4|V'^c|+8,
\end{eqnarray*}
upon using (\ref{eq:deg12}). Using in (\ref{eq:y_p2}),
\begin{eqnarray*}
\mathcal{Y}(G') &\ge& 5|V'^b|-4|V'^c|+8 - 4(3|V'^c|-6),\\
&=&5|V'^b|-16|V'^c|+32.  
\end{eqnarray*}
We see that $\mathcal{Y}(G') >0$, whenever $R=1-\frac{|V'^c|}{|V'^b|} \ge \frac{11}{16}$.
\end{IEEEproof}
By Lemma \ref{lemma:1}, we have shown that there is a codeword-supporting edge in $\mathcal{C}^\dagger(G')$.
\subsubsection{Codeword-supporting subgraph for $G$}
\begin{figure}[ht!]
\centering
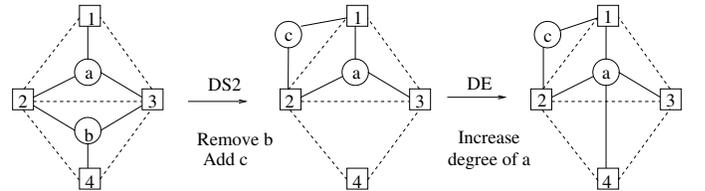
\caption{Illustration of DS and DE operations.}
\label{fig:operations}
\end{figure}

To extend the proof to a general planar Tanner graph, we show that a series of simple modifications can transform the check inverse $G'$ to the original Tanner graph $G$. We begin by defining three basic operations on a generic planar Tanner graph $P$.
\begin{enumerate}
\item {\it DS1:} Remove a degree-3 bit node in $P$ and add a degree-1 bit node to some check node.
\item {\it DS2:} Remove a degree-3 bit node in $P$ and add a degree-2 bit node to a pair of check nodes keeping the resulting graph planar.
\item {\it DE:} Increase the degree of a degree-3 bit node by connecting it to one or more check nodes so that the resulting graph is still planar. The resulting increase in degree is called the {\it expansion factor} of DE.
\end{enumerate}
The abbreviation DS stands for Degree Shrinking, and DE stands for Degree Expansion. These operations are illustrated in Fig \ref{fig:operations}.
In Fig. \ref{fig:operations}, the solid lines are the edges of the Tanner graph and the dotted lines are edges of the check graph. Note that the check graph is unaltered by the DS and DE operations.

\begin{prop}
Let $G$ be a planar Tanner graph with check inverse $G'$. Then, $G$ can be obtained from $G'$ by a series of DS1, DS2 and DE operations.
\label{prop:ds}
\end{prop}
\begin{IEEEproof}[Sketch of proof]
In the process of constructing $G'$, the following observations can be made: (1) degree-3 nodes of $G$ are retained in $G'$, (2) some degree-2 and degree-1 nodes may be dropped, and (3) higher degree ($\ge4$) nodes of $G$ result in multiple degree-3 nodes in $G'$. 
 
The operations DS1 and DS2 restore the dropped degree-2 and degree-1 nodes, while a following DE operation creates higher degree nodes. Note that DS1 and DS2 create "empty" faces in the check graph of $G'$ while DE makes a set of faces correspond to a single bit node of higher degree.

Also note that if we start with $G'$, DE by a factor of $x$ is always preceded by $x$ DS operations since $x$ empty faces should be created before DE in order to preserve planarity. These operations recursively position degree 2 and degree 1 bit nodes to the positions as in $G$ and also create bit nodes of higher degree matching to those in $G$.
\end{IEEEproof}

\begin{example}
Consider the Tanner graph $G$ and its check inverse $G'$ in Fig. \ref{fig:chkinv} of Example \ref{ex:GtoG'}. Starting with $G'$ we can obtain $G$ through a series of DS1, DS2 and DE operations. This is illustrated in Fig. \ref{fig:inv_to_G}. Note that $\mathcal{C}(G)$ remains a valid check graph of the resulting graph after every operation.
\begin{figure}[ht!]
\centering
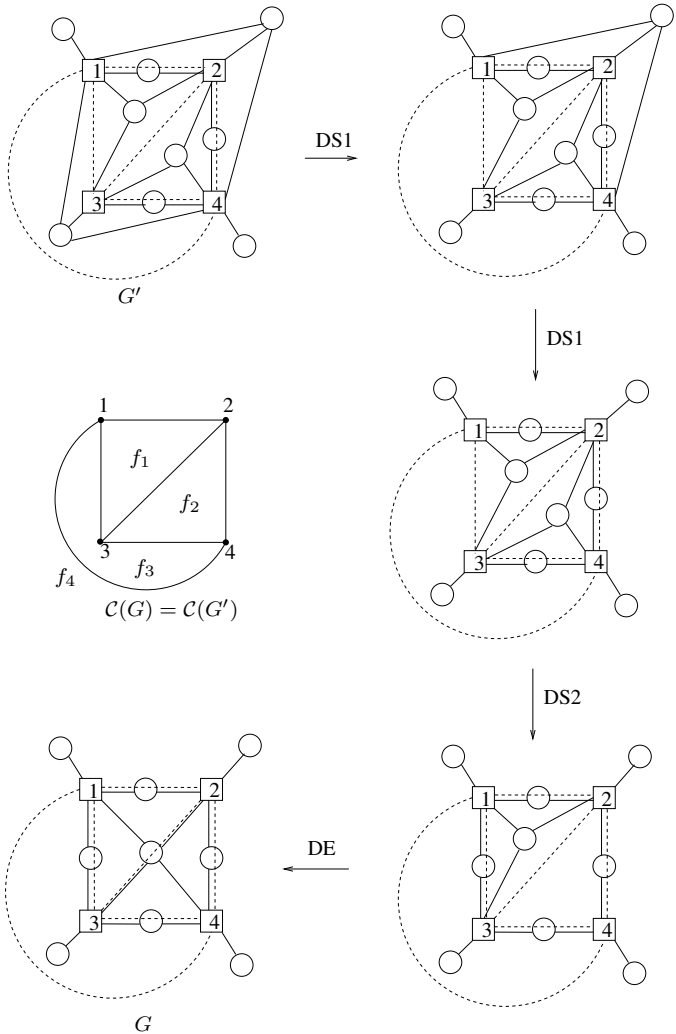
\caption{Illustration of Proposition \ref{prop:ds}.}
\label{fig:inv_to_G}
\end{figure}
\end{example}

We are now ready to prove the existence of a codeword-supporting edge in $\mathcal{C}^\dagger(G)$. The approach is to show that the operations DS1, DS2 and DE cannot decrease $\mathcal{Y}(G')$. Hence, at the end of the necessary series of DS and DE operations to get $G$ from $G'$, we have $\mathcal{Y}(G)>0$. The result is proved in the following lemma.
\begin{lem}
Let $G$ be a planar Tanner graph representing a code with rate $R\ge11/16$. Then there is an edge in $\mathcal{C}^\dagger(G)$ that is codeword-supporting. 
\label{lemma:Gedge}
\end{lem}
\begin{IEEEproof}
We will prove by showing that the operations DS and DE on the bit nodes of $G'$ to get $G$ are such that $\mathcal{Y}(G)\ge \mathcal{Y}(G')$ (see (\ref{eq:y_p2}) for definition). Since both $\mathcal{C}^\dagger(G')$ and $\mathcal{C}^\dagger(G)$ have same structure when seen as graphs, the term $\sum_{e^\dagger \in E^\dagger(G')} |V^c_{e^\dagger}|=\sum_{e^\dagger \in E^\dagger(G)} |V^c_{e^\dagger}|$. The change will be in $\sum_{e^\dagger \in E^\dagger(G')} wt(e^\dagger)$. Let 
$$\Delta=\sum_{e^\dagger \in E^\dagger(G)} wt(e^\dagger)-\sum_{e^\dagger \in E^\dagger(G')} wt(e^\dagger).$$ 
We will show that $\Delta\ge0$ to claim the lemma.

Let $H$ be the planar graph obtained at some intermediate step in the transformation from $G'$ to $G$. Let us see how DS and DE operations affect $\sum_{e^\dagger \in E^\dagger(H)} wt(e^\dagger)$.

The operation DS2 reduces the number of degree-3 bit nodes by one, and increases the number of degree-2 bit nodes by one. Let $\eta_2^*$ and $\eta_3^*$ be the new values of the terms $\eta_2$ and $\eta_3$ in (\ref{eta}) after the operation DS2. Let $\delta_{DS2}$  be the change in  $\sum_{e^\dagger \in E^\dagger(H)} wt(e^\dagger)$ when a DS2 is performed. We see that
\begin{align*}
\eta_3^* &= \eta_3 - 3\\
\eta_2^* &= \eta_2 + 5\\
\delta_{DS2} &= (\eta_3^* + \eta_2^*) - (\eta_3 + \eta_2) = 2 >0
\end{align*}

Since DE by a factor of $x$ is preceded by $x$ DS operations, we will study the net effect.  DE with expansion factor of $x$ preceded by $x$ DS2s results in the following:  
\begin{enumerate}
\item[(i)] reduces the number of degree-3 bit nodes by $x+1$
\item[(ii)]  increases the number of degree-2 bit nodes by $x$ 
\item[(iii)] introduces a bit node of degree $x+3$.
\end{enumerate}
Effect of (i) and (ii) can be readily derived as before. As (iii) involves introduction of a bit node, it increases $\mathcal{Y}(H)$ by a positive quantity, say $\alpha_x$. Let $\delta_{DE}$  be the change in  $\sum_{e^\dagger \in E^\dagger(H)} wt(e^\dagger)$ when a DE is performed. Therefore,
\begin{equation}
\delta_{DE} = -3(x+1) + 5(x) + \alpha_x.
\label{delta_p2}
\end{equation}
Since $\delta_{DE}$ is non-negative for $x>1$, it is enough to compute $\alpha_1$ and show that $\delta_{DE}$ is non-negative for $x=1$. When the expansion factor is one, a degree-3 bit node $v_b$ becomes a degree-4 bit node, and $v_b$ is identified with two faces of the new check graph having a common edge. This is equivalent to saying that $v_b$ is identified with an edge in the dual of the check graph. Therefore, $\alpha_1 = 1$. Substituting in (\ref{delta_p2}), we get $\delta_{DE} =0$ for $x=1$.

As $q(v_b)$ of a degree-1 bit node $v_b$ is more than $q(v'_b)$ of a degree-2 bit node $v'_b$, $\delta_{DS1}$ is non-negative whenever DS1 is used in place of DS2 where $\delta_{DS1}$ is the change in $\sum_{e^\dagger \in E^\dagger(H)} wt(e^\dagger)$ when a DS1 is performed.

Since $G$ can be obtained from a series of DS1, DS2 and DE operations, and since $\delta_{DS1}$, $\delta_{DS2}$, and $\delta_{DE}$ are all non-negative, $\Delta \ge 0$. Hence, the lemma is proved.
\end{IEEEproof}

\subsubsection{Proof of Theorem \ref{thm:weak}}
By Lemma \ref{lemma:Gedge}, $\mathcal{C}^\dagger(G)$ has an edge $e^\dagger$ such that $|\mathcal{I}(V^c_{e^\dagger}|)>|V^c_{e^\dagger}|$. We see that $|V^c_{e^\dagger}|=4$ by Proposition \ref{prop:vcu}. Hence, by Proposition \ref{prop:dmin}, the code defined by $G$ has a codeword of weight at most $|V^c_{e^\dagger}| +1 =5$. This proves Theorem \ref{thm:weak}.

\section{Proof of Main Result}
\label{sec:proof-main-result}
We now provide the proof of Theorem \ref{thm:main}. The method and steps of proof are similar to that of the proof of the weaker Theorem \ref{thm:weak}. A codeword-supporting subgraph will be shown to exist in the dual of check graph of check inverse by a similar counting argument. The result will then be extended by DS and DE operations to the original graph. 

The main change is that an edge of the dual need not be codeword-supporting for lower rates. We will show that among the neighborhoods of vertices of the dual of check graph with $p$ nodes (for a suitably chosen $p$), there exists a codeword-supporting subgraph, which provides a bound on minimum distance. 

We will impose a girth condition on the dual of check graph, so that the neighborhoods become trees for small $p$. Since the dual of check graph is regular with degree 3, the vertex neighborhoods will be rooted trees with mostly degree-3 nodes except near the leaves. The girth condition on the dual of the check graph will be later shown to translate into a condition on the rate of the code defined by the original Tanner graph. 

The type of neighborhood structure in the dual check graph is captured by 3-trees defined below.

\subsection{3-trees}
A {\it 3-tree} rooted at a vertex $v_r$ is a rooted tree in  which the root $v_r$ has at most 3 children and all other nodes have at most 2 children. The {\it depth} of a vertex $v$ is the length of the path from the root $v_r$ to the vertex $v$. The set of all nodes at a given depth is called a {\it level} of the tree. The root node is at depth zero. The {\it height} of a tree is the length of the path from the root to the deepest node in the tree. A {\it complete} 3-tree is one in which every level, except possibly the last, is completely filled.

\begin{figure}[ht!]
\centering
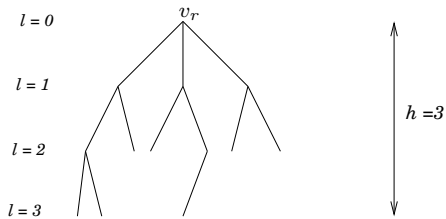
\caption{A Complete 3-tree rooted at $v_r$.}
\label{fig:3tree}
\end{figure}

The number of nodes at level $l$ (except possibly the last) of a complete 3-tree is $3.2^{l-1}$ and number of nodes up to and including level $l$ is $3.2^l - 2$. Fig. \ref{fig:3tree} shows a picture of a complete 3-tree.

Here are a few results on complete 3-trees.
\begin{enumerate}
\item Let $p$ be the number of vertices of a complete 3-tree. Then $p=3.2^{l(p)-1} - 2 + z(p)$ for a suitable level $l(p)$ such that $0 \le z(p) < 3.2^{l(p)-1}$, where $z(p)$ denotes the number of nodes in the last level. 
\item Height of the complete 3-tree with $p$ vertices is 
$$h(p) = \begin{cases} 
  l(p)\;\;\;\;\;\;, z(p) \neq 0\\
  l(p)-1\;\;,z(p)=0
\end{cases}.$$
\item There are $t(p)= \binom{3.2^{l(p)-1}}{z(p)}$ complete 3-trees of height $h(p)$ rooted at a given vertex $v$. Each such tree is called a {\it realization} of the complete 3-tree rooted at $v$. 
\item There are three branches from the root of a complete 3-tree on $p$ nodes. The number of nodes up to level $l<h(p)$ on one branch (excluding the root node) is $\frac{1}{3}(3.2^l - 3)=2^l-1$.
\end{enumerate}

The parameters $l(p)$, $z(p)$, $h(p)$ and $t(p)$ are evaluated for some values of $p$ in Table \ref{Table:pvals}.
\begin{table}[ht]
\centering 
\begin{tabular}{|c|c|c|c|c|}
 	\hline
$p$ &  $l(p)$    &   $z(p)$  & $h(p)$   & $t(p)$  \\
\hline
\hline
2 & 1 & 1 & 1 & 3 \\
\hline
3 & 1 & 2 & 1 & 3 \\
\hline
4 & 2 & 0 & 1 & 1 \\
\hline
5 & 2 & 1 & 2 & 6 \\
\hline
8 & 2 & 4 & 2 & 15 \\
\hline
10 & 3 & 0 & 2 & 1 \\
\hline
\end{tabular}
\caption{Parameters of a complete 3-tree for some values of $p$}
\label{Table:pvals}
\end{table}

\subsection{Complete 3-graphs in $\mathcal{C}^\dagger(G)$}
Since $\mathcal{C}^\dagger(G)$ is planar with uniform degree 3, we can look for complete 3-trees on $p$ vertices in the vertex neighbourhoods of $\mathcal{C}^\dagger(G)$. Given $p$,  each root has $t(p)$ complete 3-trees of height $h(p)$ only when the  girth $g$ of $\mathcal{C}^\dagger(G)$ satisfies 
\begin{equation}
  \label{eq:girth}
g \ge 2h(p)+1,   
\end{equation}
where $h(p)$ is the height defined as before.

Let $V^\dagger(G)$ be the vertex set of $\mathcal{C}^\dagger(G)$. Let $V_{i,j}^\dagger \subseteq V^\dagger(G)$ be the vertex set of the $j$-th realization of a complete 3-subtree rooted at node $v_i$ with $p$ vertices. Let the subgraph induced by $V_{i,j}^\dagger$ be denoted as $\mathcal{C}_{i,j}^\dagger(G)$ for simplicity. These induced subgraphs from now on will be referred to as {\it complete 3-graphs}. 

Note that $\mathcal{C}_{i,j}^\dagger(G)$ need not be a tree due to the presence of one or more extra edges that can create cycles. However, under the girth condition, these additional edges can only be between leaves of the complete 3-tree creating cycles of length $2h(p)+1$. These additional edges between the leaves of a realization of a complete 3-tree in the neighborhood of a node are said to be {\it cycle-creating} with respect to that particular realization. 

The number of realizations with root $v_i$ in which an edge $e^\dagger$ is cycle-creating is called the {\it recurrence  number} of the edge $e^\dagger$ with respect to the vertex $v_i$, and is denoted as $r_{v_i}(e^\dagger)$. The total number of all realizations with respect to all roots in which  $e^\dagger$ is cycle-creating is called the {\it total recurrence number} of the edge and denoted $r(e^\dagger)$. We readily see that
\begin{equation}
  \label{eq:rec}
   r(e^\dagger) = \sum_{v_i \in V^\dagger(G)} r_{v_i}(e^\dagger).  
\end{equation}
An edge that is cycle-creating for at least one realization (or edge with positive total recurrence number) is called a {\it cycle-edge}. The set of all cycle-edges in $\mathcal{C}^\dagger(G)$ is denoted by $\mathfrak{B}(G)$.

\begin{example}
Consider the dual graph $\mathcal{C}^\dagger(G)$ shown in Fig. \ref{fig:com3gph_ex}. 
\begin{figure}[ht!]
\centering
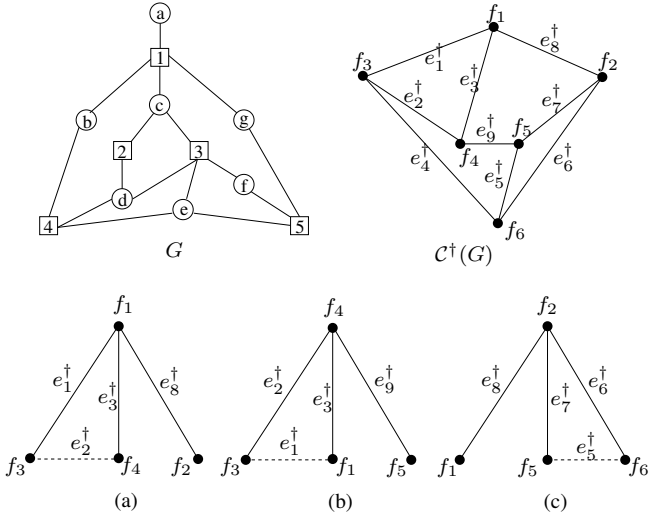
\caption{An illustration of complete 3-graphs rooted at few nodes for $p=4$.}
\label{fig:com3gph_ex}
\end{figure}
We will show some complete 3-graphs in $\mathcal{C}^\dagger(G)$ of Fig. \ref{fig:com3gph_ex} for $p=4$. Fig. \ref{fig:com3gph_ex}(a) shows a complete 3-graph rooted at $f_1$ in which $e_2^\dagger$ is a cycle-creating edge. Similarly, \ref{fig:com3gph_ex}(b),(c) shows complete 3-graphs rooted at $f_4$ and $f_2$ respectively. The corresponding cycle creating edges are shown in dashed lines. Observe that $e_2^\dagger$ which is cycle-creating for the complete 3-graph rooted at $f_1$, but is not cycle-creating for the complete 3-graph rooted at $f_4$.

For the complete 3-graph rooted at $f_4$, $e_1^\dagger$ is cycle-creating. Since $p=4$, $t(p)=1$ i.e, there is only one realization per root node.  Hence, $r_{f_4}(e_1^\dagger) = 1$. Also note that $e_1^\dagger$ is not cycle-creating for other complete 3-graphs rooted at other nodes. Therefore, $r(e_1^\dagger) = 1$. Using similar arguments, we can show that $r(e_i^\dagger) =1$ for $i=\{2,3,5,6,7\}$. These edges  are cycle-edges in $\mathcal{C}^\dagger(G)$. All other edges are non-cycle-edges; hence, their recurrence number is zero.
\end{example}

Let $W_{v_i,j}=wt(\mathcal{C}_{i,j}^\dagger(G))$ be the weight of the $j$-th realization of a complete 3-graph rooted at $v_i$. Let $V^c_{i,j}$ be the set of check nodes forming the faces corresponding to the vertices $V_{i,j}^\dagger$ in $\mathcal{C}^\dagger(G)$ i.e. $V^c_{i,j}=V^c_U$ with $U=V^\dagger_{i,j}$ as in (\ref{eq:vcu}). By Proposition \ref{prop:vcu},
\begin{equation}
  \label{eq:vcij}
|V^c_{i,j}|\le p+2 - c_{v_i,j},  
\end{equation}
where $c_{v_i,j}$ is the number of cycles in the $j$-th realization of the complete 3-graph on $p$ nodes rooted at $v_i$. By definition, $\mathcal{C}_{i,j}^\dagger(G)$ is codeword-supporting if $W_{v_i,j} - |V^c_{i,j}| > 0$.

Consider the summation,
\begin{equation}
  \mathcal{Y}(G) = \sum_{v_i \in V^\dagger(G)} \sum_{j = 1}^{t(p)} \left(W_{v_i,j}-|V^c_{i,j}|\right) \label{eq:Ygen}  
\end{equation}
where $t(p)$ is the number of realizations of complete 3-trees on $p$ vertices rooted at a particular node assuming the girth condition (\ref{eq:girth}). We will show that $\mathcal{Y}(G)>0$ to establish the existence of a codeword-supporting subgraph among the complete 3-graphs of $\mathcal{C}^\dagger(G)$. Let $m=|V^c|$ and $n=|V^b|$ be the number of check nodes and bit nodes of $G$. Hence, $|V^{\dagger}(G)|=2m-4$. Using (\ref{eq:vcij}) in (\ref{eq:Ygen}), we get
\begin{align}
\mathcal{Y}(G) &\ge \sum_{v_i \in V^\dagger(G)} \sum_{j = 1}^{t(p)} W_{v_i,j} - [(2m-4)pt(p) + 2(2m-4)t(p)]\nonumber\\
&\phantom{= \sum_{v_i \in V^\dagger(G)} } + \sum_{v_i \in V^\dagger(G)} \sum_{j = 1}^{t(p)} c_{v_i,j},\label{eq:3}  
\end{align}
To show $\mathcal{Y}(G)>0$, we simplify the terms in the right hand side of (\ref{eq:3}). 
\subsection{Simplifying $\sum_{v_i \in V^\dagger(G)} \sum_{j = 1}^{t(p)} c_{v_i,j}$}
We begin by relating the number of cycles $c_{v_{i,j}}$ to  recurrence number of edges.
\begin{prop} The following equality holds:
$$\sum_{v_i \in V^\dagger(G)} \sum_{j = 1}^{t(p)} c_{v_i,j} = \sum_{e_l^\dagger \in \mathfrak{B}(G)} r(e_l^\dagger)$$
\label{prop:rec}
\end{prop}
\begin{IEEEproof}
We see that 
$$\sum_{v_i \in V^\dagger(G)} \sum_{j = 1}^{t(p)} c_{v_i,j}=\sum_{v_i \in V^\dagger(G)} \sum_{e_l^\dagger \in \mathfrak{B}(v_i)} r_{v_i}(e_l^\dagger),$$
where $\mathfrak{B}(v_i)$ is the set of edges that are cycle-creating in any of the complete 3-graphs rooted at $v_i$. Now,
\begin{align}
\sum_{v_i \in V^\dagger(G)} \sum_{j = 1}^{t(p)} c_{v_i,j}&= \sum_{e_l^\dagger \in \mathfrak{B}(G)} \sum_{v_i \in V^\dagger(G)} r_{v_i}(e_l^\dagger)\nonumber\\
       &= \sum_{e_l^\dagger \in \mathfrak{B}(G)} r(e_l^\dagger).\label{eq:1}
\end{align}
\end{IEEEproof}
\subsubsection{Occupied and unoccupied edges}
In the computation of $\mathcal{Y}(G)$, the cycle edges identified with degree-2 bit nodes should be treated separately. Such cycle edges are classified next.

Each degree-2 bit node in $G$ is identified as an edge in $\mathcal{C}(G)$. An edge $e$ in $\mathcal{C}(G)$ is said to be {\it occupied} by a degree-2 bit node if the bit node is connected to the check nodes of $e$. Else it is said to be {\it unoccupied}. Since there is a one to one correspondence between edges of $\mathcal{C}(G)$ and edges of its dual $\mathcal{C}^\dagger(G)$, we use the terms occupied and unoccupied for edges of dual as well.

Similarly, we can talk of occupied and unoccupied check nodes.  A check node of a planar Tanner graph  is said to be occupied if there is a degree-1 bit node connected to it. The degree-1 bit node is said to occupy the check node. Otherwise, the check node is said to be unoccupied.

Let $\mathfrak{B}(G)$ be the set of cycle-edges, and let $\mathfrak{B}_o(G)$ and $\mathfrak{B}_u(G)$ be the set of occupied and unoccupied cycle-edges. We see that $\mathfrak{B}_o(G)$ and $\mathfrak{B}_u(G)$ partition $\mathfrak{B}(G)$ so that the following holds: 
\begin{equation}
|\mathfrak{B}(G)| = |\mathfrak{B}_o(G)| + |\mathfrak{B}_u(G)|. \label{eq:B}
\end{equation}
\begin{example}
In the Tanner graph of Figs. \ref{fig:induced_bit} and \ref{fig:com3gph_ex}, the set of occupied edges are given by $\{e_4^\dagger,e_6^\dagger,e_7^\dagger\}$ as seen from Example \ref{ex:dual-check-graph}, and the set of cycle-edges is given by $\mathfrak{B}(G)=\{e_2^\dagger,e_3^\dagger,e_5^\dagger,e_6^\dagger,e_7^\dagger\}$. Hence, we see that $\mathfrak{B}_o(G)=\{e_6^\dagger,e_7^\dagger\}$ and $\mathfrak{B}_u(G)=\{e_2^\dagger,e_3^\dagger,e_5^\dagger\}$.
\end{example}
Using the partition of $\mathfrak{B}(G)$ in (\ref{eq:1}), we see that
\begin{equation}
  \label{eq:oc}
  \sum_{v_i \in V^\dagger(G)} \sum_{j = 1}^{t(p)} c_{v_i,j}=\sum_{e_l^\dagger \in \mathfrak{B}_o(G)} r(e_l^\dagger)+\sum_{e_l^\dagger \in \mathfrak{B}_u(G)} r(e_l^\dagger).
\end{equation}
\begin{example}
\label{ex:occup-unocc-edges}
In the illustration of the DS and DE operations in Fig. \ref{fig:inv_to_G}, we stated that the check graph of $G$ and $G'$ are the same. However, the set of occupied edges and check nodes changes because of the changes in the number of degree-2 and degree-1 nodes. In the check graph $\mathcal{C}(G')$ in Fig. \ref{fig:inv_to_G}, the set of occupied edges is $\{(1,2),(2,4),(3,4)\}$, and the set of occupied nodes is $\{1,4\}$. However, in $\mathcal{C}(G)$ in Fig. \ref{fig:inv_to_G}, the set of occupied edges is $\{(1,2),(1,3),(2,4),(3,4)\}$, and the set of occupied nodes is $\{1,2,3,4\}$. Since new degree-2 and degree-1 nodes can possibly be added in $G$ through the DS operations, some unoccupied edges and check nodes in $\mathcal{C}(G')$ become occupied in $\mathcal{C}(G)$.
\end{example}
\subsubsection{Singular nodes}
\label{sec:singular-nodes}
Another special situation arises with multiple edges in check graphs. Note that, by construction, multiple edges can arise in check graphs, if they do not result in a face enclosed by two edges. The effect of multiple edges in check graphs is characterized next.

\begin{figure}[ht!]
\centering
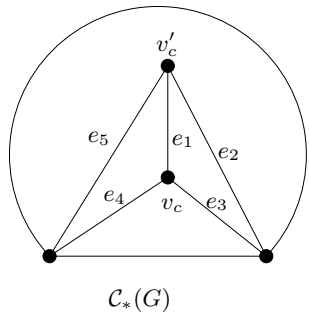
\caption{Singular nodes.}
\label{fig:sing_defn}
\end{figure}

\noindent \textit{Definition:} Let $G$ be a planar Tanner graph, and let $\mathcal{C}(G)$ be its check graph. Let $\mathcal{C}_*(G)$ be a subgraph of $\mathcal{C}(G)$ consisting of a maximal planar graph on 4 vertices plus an edge that leads to an external face of length 2 in  $\mathcal{C}_*(G)$ as shown in Fig. \ref{fig:sing_defn}.  In addition, we impose the constraint that the interior faces of $\mathcal{C}_*(G)$ are faces in $\mathcal{C}(G)$.
A degree-1 bit node in $G$ is said to be \textit{singular} if it occupies a check node in $G$ that corresponds to one of the interior nodes of $\mathcal{C}_*(G)$. In Fig. \ref{fig:sing_defn}, a degree-1 bit node that occupies $v_c$ or $v'_c$ is singular. 

The number of singular nodes in a planar Tanner graph $G$ is denoted $s_G$. The following proposition relates $s_G$ to the recurrence number.
\begin{prop}
Let $G$ be a planar Tanner graph with $s_G$ singular nodes. Let $\mathfrak{B}_u(G)$ be the set of unoccupied cycle-edges in $\mathcal{C}^\dagger(G)$ for the complete 3-graphs on 4 nodes. If
\begin{equation*}
\sum_{e_l^\dagger \in \mathfrak{B}_u(G)} r(e_l^\dagger) <   s_{G},
\end{equation*}
there exists a codeword-supporting subgraph on 4 nodes in $\mathcal{C}^\dagger(G)$.
\label{prop:sing}
\end{prop}
\begin{IEEEproof}
Let $\sum_{e_l^\dagger \in \mathfrak{B}_u(G)} r(e_l^\dagger) <   s_{G}$. Since $\sum_{e_l^\dagger \in \mathfrak{B_u}} r(e_l^\dagger)$ is non-negative $s_G$ is at least 1. Hence, there is at least one subgraph of the form $\mathcal{C}_*(G)$  as shown in Fig. \ref{fig:sing_defn} in  $\mathcal{C}(G)$. Let $\mathfrak{B}$ be the set of edges in $\mathcal{C}^\dagger(G)$ corresponding to the set of interior edges $\{e_1,e_2,e_3,e_4,e_5\}$ in $\mathcal{C}_*(G)$. The edges of $\mathfrak{B}$ are shown as dashed lines in Fig. \ref{fig:cyc_edges_sing}. Note these the edges are cycle-edges for $p=4$.

\begin{figure}[htb]
\centering
\includegraphics{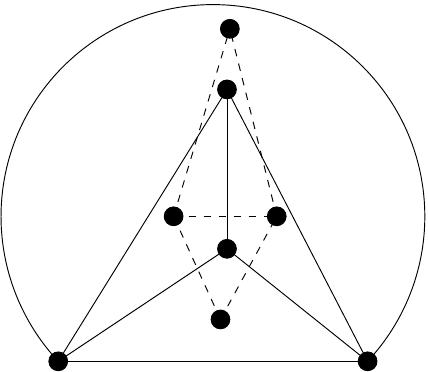}
\caption{Cycle edges in the dual for $p=4$.}
\label{fig:cyc_edges_sing}
\end{figure}

Let $\mathcal{C}_*^i(G)$, $i=1,2...k$ be the subgraphs in $\mathcal{C}(G)$ of the form $\mathcal{C}_*(G)$. Let $s^i_G$ be the number of singular nodes in $\mathcal{C}_*^i(G)$ and $\mathfrak{B}^i$ be the set of 5 cycle-edges corresponding to the interior edges in $\mathcal{C}_*^i(G)$. Observe that $\mathfrak{B}^i$ are disjoint. Let $\mathfrak{B}^i_u$ be the subset of edges of $\mathfrak{B}^i$ that are unoccupied in $\mathcal{C}^\dagger(G)$. Let $\mathfrak{B}^*_u = \bigcup_{i=1}^k \mathfrak{B}^i_u$. Then, $\mathfrak{B}^*_u \subseteq \mathfrak{B}_u(G)$. Hence,
\begin{align*}
\sum_{e_l^\dagger \in \mathfrak{B}_u^*} r(e_l^\dagger) \le \sum_{e_l^\dagger \in \mathfrak{B}_u(G)} r(e_l^\dagger) < s_G,\\
\sum_{i=1}^k \sum_{e_l^\dagger \in \mathfrak{B}_u^i} r(e_l^\dagger) < \sum_{i=1}^k s^i_G, \\
\sum_{i=1}^k \left (\sum_{e_l^\dagger \in \mathfrak{B}_u^i} r(e_l^\dagger) - s^i_G \right) < 0.
\end{align*}
Thus, there exists $j$, $1 \le j\le k$, such that,
\begin{align*}
\sum_{e_l^\dagger \in \mathfrak{B}_u^j} r(e_l^\dagger) <  s^j_G,  \\
|\mathfrak{B}_u^j| \le \sum_{e_l^\dagger \in \mathfrak{B}_u^j} r(e_l^\dagger) < s^j_G, \\
|\mathfrak{B}_u^j| < s^j_G.
\end{align*}
Thus, there are at most $s^j_G -1$ edges in $\mathfrak{B}^j$ that are unoccupied. In other words, there are at least $6-s^j_G$ occupied edges in $\mathfrak{B}^j$. Therefore, the 4 nodes in $\mathcal{C}_*^j(G)$ induce at least $6-s^j_G +s^j_G = 6$ bit nodes in $G$. Since, there are 4 faces in $\mathcal{C}_*^j(G)$, there exists a codeword supporting subgraph on 4 nodes in $\mathcal{C}^\dagger(G)$.
\end{IEEEproof}

\subsection{Minimum distance bound for $G'$}
The next lemma is a generalization of Lemma \ref{lemma:1} from edges to complete 3-graphs for the main result.
\begin{lem} Let $G$ be a planar Tanner graph with check inverse $G'$. Let $G'$ define a code with rate $R$. Then, there exists a codeword-supporting subgraph on $p$ vertices in $\mathcal{C}^\dagger(G')$ where
$$p=  \left\lceil \frac{7-8R}{2(2R-1)} \right\rceil$$
is such that the girth condition (\ref{eq:girth}) is satisfied. 
\label{lemma:3g}
\end{lem}
As before, we will prove the result by showing that $\mathcal{Y}(G')>0$. 
\subsubsection*{Simplifying $\sum_{v_i \in V^\dagger(G')} \sum_{j = 1}^{t(p)} W_{v_i,j}$}
Let us now consider the summation of weights term in (\ref{eq:3}). As done in the illustrative proof of Theorem \ref{thm:weak}, we let
$$\sum_{v_i \in V^\dagger(G')} \sum_{j = 1}^{t(p)} W_{v_i,j}=\sum_{v_b\in V'^b}q(v_b),$$
where $q(v_b)$ is the number of times a node $v_b$ of $G'$ is induced by $V^c_{i,j}$ for all $i$ and $j$. Let
\begin{equation}
  \label{eq:etadf}
\eta_i = \sum_{{v_b \in V_i^{'b}}} q(v_b),  
\end{equation}
where $V_i^{'b}$ is the set of degree-$i$ bit nodes in $G'$. Then,
\begin{align}
\sum_{v_i \in V^\dagger(G')} \sum_{j = 1}^{t(p)} W_{v_i,j} = \eta_1 + \eta_2 + \eta_3. \label{eq:wexpn}
\end{align}
We will begin with calculation of $\eta_3$. A degree-3 bit node $v_b$ in $G'$ is identified with a node $f$ in $\mathcal{C}^\dagger(G')$. So, $q(v_b)$ is same as the number of complete 3-graphs that contain $f$, which equals $pt(p)$. Hence, we see that
$$q(v_b) = pt(p) \text{ for }v_b \in V^{'b}_3$$
Therefore,
\begin{equation}
\eta_3 = (2m-4)pt(p). \label{q3}
\end{equation}
The computation of $\eta_2$ and $\eta_1$, as shown in Appendices \ref{sec:eta_2} and \ref{sec:eta_1}, results in the following.
\begin{align}
\eta_2 & = |V'^b_2|(\frac {4} {3} p + \frac {2}{3})t(p) - \sum_{e_l^\dagger \in \mathfrak{B}_o(G')}  r(e_l^\dagger) \label{q2}\\
\eta_1 & \ge |V'^b_1|(\frac {4} {3} p + \frac {2}{3})t(p) - s_{G'}(p), \label{q1}
\end{align}
where $s_{G'}(p)$ is given by,
\begin{equation}
s_{G'}(p) = \begin{cases} 
  s_{G'} \;\;\;\;\;\;\text{when~}  p = 4\\
  0 \;\;\;\;\;\;\text{else}
\end{cases}
\label{s(p)}
\end{equation}
with $s_{G'}$ being the number of \textit{singular nodes} in $G'$ as discussed in Section \ref{sec:singular-nodes}.

Using (\ref{q3}), (\ref{q2}) and (\ref{q1}), we get
\begin{multline*}
\sum_{v_i \in V^\dagger(G')} \sum_{j = 1}^{t(p)} W_{v_i,j}\ge(2m-4)pt(p)\\
+|V'^b_2|(\frac {4} {3} p + \frac {2}{3})t(p)+|V'^b_1|(\frac {4} {3} p + \frac {2}{3})t(p)\\
-\sum_{e_l^\dagger \in \mathfrak{B}_o(G')}  r(e_l^\dagger)- s_{G'}(p).
\end{multline*}
Using $|V'^b_1|  + |V'^b_2| = n- (2m-4)$, we get
\begin{multline}
  \label{eq:wsimp}
  \sum_{v_i \in V^\dagger(G')} \sum_{j = 1}^{t(p)} W_{v_i,j}\ge[\frac{4}{3}pn+\frac{2}{3}n-\frac{1}{3}p(2m-4)-\frac{2}{3}(2m-4)]t(p)\\
-\sum_{e_l^\dagger \in \mathfrak{B}_o(G')}  r(e_l^\dagger)- s_{G'}(p).
\end{multline}
\begin{IEEEproof}[Proof of Lemma \ref{lemma:3g}]
Using (\ref{eq:oc}) and (\ref{eq:wsimp}) in the expression for $\mathcal{Y}(G')$ in (\ref{eq:3}), we get 
\begin{align}
\mathcal{Y}(G') & \ge  (\frac{4}{3}pn +\frac{2}{3}n - \frac{4}{3} p (2m-4) - \frac{8}{3} (2m-4))t(p)  \nonumber \\
&\phantom{\frac{4}{3}pn +}+\sum_{e_l^\dagger \in \mathfrak{B}_u(G')} r(e_l^\dagger) - s_{G'}(p).  \label{X}
\end{align}
Let $X = (\frac{4}{3}pn +\frac{2}{3}n - \frac{4}{3} p (2m-4) - \frac{8}{3} (2m-4))t(p) $. Hence,
\begin{align}
\mathcal{Y}(G') & \ge X + \sum_{e_l^\dagger \in \mathfrak{B}_u(G')} r(e_l^\dagger) - s_{G'}(p) \label{eq:yexpn_x}
\end{align}

We fix $p$ to be the smallest integer that results in $X > 0$. This is readily seen to be
\begin{equation}
p = \left \lceil \frac{7-8R}{2(2R-1)} \right \rceil. \label{pexpn}
\end{equation}

\noindent \textit{Case} 1: $p \ne 4$.\\
When $p$ given by (\ref{pexpn}) is not equal to 4, $s_{G'}(p)=0$ by (\ref{s(p)}) and hence the term $\sum_{e_l^\dagger \in \mathfrak{B}_u(G')} r(e_l^\dagger) - s_{G'}(p)$ is non-negative. This implies that, for this $p$, $\mathcal{Y}(G') > 0$  and there exists a codeword-supporting complete 3-graph on $p = \left \lceil \frac{7-8R}{2(2R-1)} \right \rceil$ nodes.

\noindent\textit{Case} 2: $p = 4$.\\
If $\sum_{e_l^\dagger \in \mathfrak{B}_u(G')} r(e_l^\dagger) - s_{G'}(p) \ge 0$, $\mathcal{Y}(G') > 0$ for $p=4$ and there exists a codeword-supporting complete 3-graph on $p = \left \lceil \frac{7-8R}{2(2R-1)} \right \rceil = 4$ nodes. Even otherwise, by Proposition \ref{prop:sing}, there exists a codeword supporting subgraph on $4$ nodes.
\end{IEEEproof}

\subsection{Minimum distance bound for $G$}
The bound for $G$ is obtained as in the proof of Theorem \ref{thm:weak} by showing that the operations DS and DE do not reduce the value of $\mathcal{Y}(G')$. The graph $G$ is obtained from $G'$ through a series of recursive operations as shown in Proposition \ref{prop:ds}.  We will see how the DS and DE operations involved in transforming $G'$ to $G$ affect the summation $\mathcal{Y}(G')$. Since both $\mathcal{C}^\dagger(G')$ and $\mathcal{C}^\dagger(G)$ have same structure when seen as graphs, the term $\sum_{v_i \in V^\dagger(G)} \sum_{j = 1}^{t(p)} |V^c_{i,j}|=\sum_{v_i \in V^\dagger(G')} \sum_{j = 1}^{t(p)} |V^c_{i,j}|$. The only change will be in $\sum_{v_i \in V^\dagger(G')} \sum_{j = 1}^{t(p)} W_{v_i,j}$. 

Let $H$ be a planar graph obtained at some intermediate step in the transformation from $G'$ to $G$. Since $\mathcal{C}(G')$ is a check graph for $H$, the girth condition is satisfied by $C^{\dagger}(H)$. We write 
\begin{equation}
  \label{eq:wh}
W(H)=\sum_{v_i \in V^\dagger(H)} \sum_{j = 1}^{t(p)} W_{v_i,j} = \eta_1(H) + \eta_2(H) + \eta_3(H),   
\end{equation}
where $\eta_i(H) = \sum_{{v_b \in V_i^{b}}} q(v_b)$ with $V_i^{b}$ being the set of degree-$i$ bit nodes in $H$.

\subsubsection{Effect of DS2}
The graph obtained after the DS1 operation on $H$ is denoted $\text{DS1}[H]$. Let $\Delta_{DS1}=W(\text{DS1}[H])-W(H)$ be the change in the weight summations because of the DS1 operation. Similar notation is used for the DS2 and DE operations.

The operation DS2 reduces the number of degree-three bit nodes by one and increases the number of degree-2 bit nodes by one. From (\ref{q3}) and (\ref{q2}),
\begin{align*}
\eta_3(\text{DS2}[H]) &= \eta_3(H) - pt(p),\\
\eta_2(\text{DS2}[H]) &= \eta_2(H) + \left(\frac{4}{3}p+\frac{2}{3} \right )t(p) - r(e^\dagger),
\end{align*}
where $e^\dagger$ is the edge in $\mathcal{C}^\dagger(\text{DS2}[H])$ identified with the new degree-2 bit node. Hence,
\begin{align} 
\Delta_{DS2} = \delta - r(e^\dagger), \label{eq:del_s2}
\end{align}
where $\delta = \frac{1}{3}p.t(p) + \frac{2}{3} t(p)>0$.

\subsubsection{Effect of DS1}
The operation DS1 reduces the number of degree-3 bit nodes by one and increases the number of degree-1 bit nodes by one. From Appendix \ref{sec:eta_1}, $q(v_b)$ for a degree-1 bit node $v_b$ is at least $\left (\frac{4}{3}p+\frac{2}{3} \right )t(p)$ except for the case when $p=4$ and $v_b$ singular. Following calculations as before, we can show that
\begin{equation}
\Delta_{DS1} = \delta - s'(p),
\label{eq:del_s1}
\end{equation}
where $s'(p) = \begin{cases} 1, \; p = 4\text{ and }v_b\text{ singular},\\
0, \;\text{else}.
\end{cases}$

\subsubsection{Effect of DE}
Let $\text{DE}(x)$ denote the DE operation by an expansion factor of $x$, which is necessarily preceded by $x$ DS operations. The operation DE$(x)$ results in the following:
\begin{enumerate}
\item[(i)] reduces the number of degree-3 bit nodes by $x+1$.
\item[(ii)]  increases the number of bit nodes of degree $\le 2$ by $x$. 
\item[(iii)]  introduces a bit node of degree $x+3$.
\end{enumerate}
Let $\mathcal{E}_*^\dagger$ be the subset of edges of $\mathcal{C}^\dagger(H)$ identified with the new degree-2 bit nodes, and let $s^+$ be the number of new degree-1 bit nodes that are singular in $\text{DE}(x)[H]$. Effect of Steps (i) and (ii) can be readily derived as before. The introduction of a bit node in Step (iii) increases $W(H)$ by a positive quantity denoted $\alpha_x$. Hence, we see that
\begin{multline*}
\Delta_{DE(x)} =  x \left (\frac{1}{3}pt(p) + \frac{2}{3}t(p) \right) - p.t(p) \\
 - \sum_{e^\dagger \in \mathcal{B}_*^{\dagger}}r(e^\dagger)  - s^+(p) + \alpha_x, 
\end{multline*}
where 
$$s^+(p) = \begin{cases} 
  s^+, \;  p = 4,\\
  0, \;\text{else},\end{cases}$$  
and $\mathcal{B}_*^\dagger$ is the set cycle edges in $\mathcal{E}_*^\dagger$. In the above equation, we use $\mathcal{B}_*^\dagger$ for the summation term instead of $\mathcal{E}_*^\dagger$ as recurrence number of a non cycle-edge is zero.

Let $\delta_{DE}(x) = x \left (\frac{1}{3}pt(p) + \frac{2}{3}t(p) \right) - pt(p) + \alpha_x $. Then,
\begin{equation}
\Delta_{DE(x)} = \delta_{DE}(x) - \sum_{e^\dagger \in \mathcal{E}_*^\dagger}r(e^\dagger) - s^+(p). \label{eq:del_de2}
\end{equation}
Since $\alpha_x$ is positive, $\delta_{DE}(x)$ is positive for $x\ge 3$. As shown in Appendix \ref{sec:gx}, 
$$\alpha_1 \ge \left (\frac{2}{3}p - \frac{2}{3}\right)t(p).$$
Substituting  $\alpha_1$ in $\delta_{DE}(x)$ for $x=1$ gives $\delta_{DE}(1) \ge 0$. As shown in Appendix \ref{sec:gx}, 
$$\alpha_2  \ge \left (\frac{1}{2}p - \frac{1}{6}z(p) - 1 \right)t(p).$$
Substituting $\alpha_2$ in  $\delta_{DE}(x)$ for $x=2$, we get
$$\delta_{DE}(2) \ge \left ( \frac{1}{3} - \frac{1}{6}z(p) + \frac{1}{6}p \right )t(p) > 0,$$
since $p > z(p)$. Hence $\delta_{DE}(x) \ge 0$ for all $x$.

\subsubsection{Codeword-supporting complete 3-graph}
We now present a generalization of Lemma \ref{lemma:Gedge} for complete 3-graphs.
\begin{lem} 
Let $G$ be a planar Tanner graph defining a code of rate $R$. Then, there exists a codeword-supporting complete 3-graph on $p = \left \lceil \frac{7-8R}{2(2R-1)} \right \rceil$ nodes in $\mathcal{C}^\dagger(G)$ provided the girth condition is satisfied.
\label{lem:G3tree}
\end{lem}
\begin{IEEEproof} We will prove the lemma by showing that $\mathcal{Y}(G) > 0$ for $p = \left \lceil \frac{7-8R}{2(2R-1)} \right \rceil$.

Let $n_{DS1}$ and $n_{DS2}$ be the number of $DS1$ and $DS2$ operations (excluding those  DS1 and DS2 operations that are performed to create empty faces before DEs) performed in obtaining $G$ from $G'$. Let $n_{DE}(x)$ be the number of DE operations with expansion factor $x$. Let $\mathcal{E}_*^\dagger$ be the subset of edges of $\mathcal{C}^\dagger(G')$ identified with the new degree-2 bit nodes resulting from these operations (see Example \ref{ex:occup-unocc-edges}). Let $s^*$ be the number of singular nodes in $G$ that are not present in $G'$. Then,
\begin{multline*}
\mathcal{Y}(G) = \mathcal{Y}(G') +   n_{DS1}\delta +  n_{DS2}\delta \\
                +\sum_{x}(n_{DE(x)}\delta_{DE}(x)) - \sum_{e^\dagger \in \mathfrak{B}_*^\dagger}r(e^\dagger) - s^*(p),
\end{multline*}
where $B_*^\dagger$ is the set of cycle edges in $\mathcal{E}_*^\dagger$ and 
$$s^*(p) = \begin{cases}
s^*,\; p=4, \\
0,\;\text{else.}
\end{cases}$$
Let $\Delta = n_{DS1}\delta +  n_{DS2}\delta +  \sum_{x}(n_{DE(x)}\delta_{DE}(x))$. From the previous section, we see that $\Delta \ge 0$. Substituting the expression for $\mathcal{Y}(G')$ from (\ref{eq:yexpn_x}), we get,
\begin{multline*}
\mathcal{Y}(G) \ge X + \Delta + \sum_{e_l^\dagger \in \mathfrak{B}_u(G')} r(e_l^\dagger) -  \sum_{e^\dagger \in \mathfrak{B}_*^\dagger}r(e^\dagger) \\
- (s_{G'}(p) + s^*(p)).
\end{multline*}
We assume that the edges in $\mathcal{E}_*^\dagger$ are unoccupied in $\mathcal{C}^\dagger(G')$. Otherwise, there will be two or more degree-2 bit nodes in $G$ identified with the same edge and $d = 2$. Therefore,
$$
\sum_{e_l^\dagger \in \mathfrak{B}_u(G')} r(e_l^\dagger) - \sum_{e^\dagger \in \mathfrak{B}_*^\dagger}r(e^\dagger)  = \sum_{e_l^\dagger \in \mathfrak{B}_u(G)} r(e_l^\dagger),
$$ 
where $\mathfrak{B}_u(G)$ is the set of unoccupied cycle-edges in $C^\dagger(G)$. Similarly, we assume that the check nodes  to which the new degree-1 singular nodes are connected to are unoccupied in $G'$ to avoid $d=2$. Hence, if $s_G$ is the number of singular nodes in $G$, we have
$$s_{G'}(p) + s^*(p) = s_G(p),$$
where $$s_G(p) = \begin{cases}
s_G, ~~p = 4 \\
0,   \;\text{~~else}.
\end{cases}
$$
Substituting the above expressions in $\mathcal{Y}(G)$, we get,
$$\mathcal{Y}(G) \ge X + \Delta + \sum_{e_l^\dagger \in \mathfrak{B}_u(G)} r(e_l^\dagger) - s_G(p).$$
From the proof of Lemma \ref{lemma:3g} we see that, $X>0$ for $p = \left \lceil \frac{7-8R}{2(2R-1)} \right \rceil$.

\noindent \textit{Case 1: $p \ne 4$}\\
\indent The term $\sum_{e_l^\dagger \in \mathfrak{B}_u(G)} r(e_l^\dagger) - s_G(p)$ is non-negative. Since $\Delta  \ge 0$, $\mathcal{Y}(G) > 0$.

\noindent\textit{Case 2: $p = 4$}\\
\indent If $\sum_{e_l^\dagger \in \mathfrak{B}_u(G)} r(e_l^\dagger) - s_{G}(p) \ge 0$, $\mathcal{Y}(G) > 0$ for $p=4$ and there exists a codeword-supporting complete 3-graph on $p = \left \lceil \frac{7-8R}{2(2R-1)} \right \rceil = 4$ nodes. Even otherwise, by Proposition \ref{prop:sing}, there exists a codeword supporting subgraph on $4$ nodes.
\end{IEEEproof}

\subsection{Girth condition and final proof}
\label{sec:girth-condition}
We now complete the proof of the main result by showing that the girth condition holds for suitable rates. 
\begin{prop}
Consider a planar Tanner graph $G$ with minimum distance $d>2$. Let $g$ be the girth of $\mathcal{C}^\dagger (G)$. Then, $g \ge 3$. Hence, the girth condition is satisfied for $p\le 4$. 
\label{prop:girth-condition-1}
\end{prop}
\begin{IEEEproof}
It is easy to see that there cannot be any loops in $\mathcal{C}^{\dagger}(G)$. We will show that multiple edges in $\mathcal{C}^{\dagger}(G)$ will result in $d=2$ to prove the proposition. Let $f_1$ and $f_2$ be two nodes in $C^{\dagger}(G)$ connected by two edges. An edge between $f_1$ and $f_2$ in the dual corresponds to a common edge between the two faces $f_1$ and $f_2$ in the $\mathcal{C}(G)$. Hence, the two edges connecting $f_1$ and $f_2$ correspond to two common edges between the faces $f_1$ and $f_2$. Since each face is of length 3 in $\mathcal{C}(G)$, both faces $f_1$ and $f_2$ have the same vertex set. Such a situation arises in the construction of $\mathcal{C}(G)$ only when two degree-3 bit nodes in $G$ have the same set of neighboring check nodes i.e. when $d=2$. Hence, when $d>2$ there are no multiple edges in $\mathcal{C}^{\dagger}(G)$.

\indent Since $p \le 4$ results in $h(p) \le 1$, the girth condition  for $p \le 4$ needs $g \ge 3$, which is satisfied for $\mathcal{C}^\dagger (G)$.
\end{IEEEproof}

\begin{example}
A situation with multiple edges is shown in Fig. \ref{fig:muledge} where $U = \{f_1, f_2\}$. 
\begin{figure}[ht!]
\centering
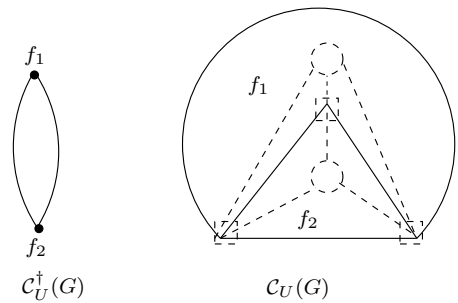
\caption{Multiple edges in the dual of check graph.}
\label{fig:muledge}
\end{figure}
In Fig. \ref{fig:muledge}, the subgraph $\mathcal{C}_U(G)$ of the check graph corresponding to multiple edges in $\mathcal{C}^\dagger(G)$ is shown. The dotted lines show the edges and nodes of $G$. We see that multiple edges in $\mathcal{C}^\dagger (G)$ result from two degree-3 bit nodes in $G$ having the same set of neighbouring check nodes, in which case $d=2$. 
\end{example}

Observe that $p =   \left\lceil \frac{7-8R}{2(2R-1)} \right\rceil \ge 0$. When $\frac{5}{8} \le  R < \frac{7}{8}$, we have $1\le p \le 4$. By Proposition \ref{prop:girth-condition-1} and Lemma \ref{lemma:3g}, there exists a codeword-supporting subgraph on $\left \lceil \frac{7-8R}{2(2R-1)} \right \rceil$ nodes. By Proposition \ref{prop:cu},
\begin{align}
d \le  \left\lceil \frac{7-8R}{2(2R-1)} \right\rceil +3. \label{eq:dmin}
\end{align}

When $R \ge \frac{7}{8}$, by Proposition \ref{prop:prelude}, $d' \le 3$ for the minimum distance $d'$ of the code defined by the check inverse $G'$. Since the DS and DE operations in the conversion from $G'$ to $G$ cannot decrease the sum $\sum_iw_i$ in Proposition \ref{prop:prelude}, the same bound holds for the minimum distance of $G$. So, we have $d\le 3= \left \lceil \frac{7-8R}{2(2R-1)} \right \rceil + 3$ for $R\ge\frac{7}{8}$. This concludes the proof of Theorem \ref{thm:main}, which is the main result of this paper. 

\begin{figure}[ht!]
  \centering
  \includegraphics[width=3.7in]{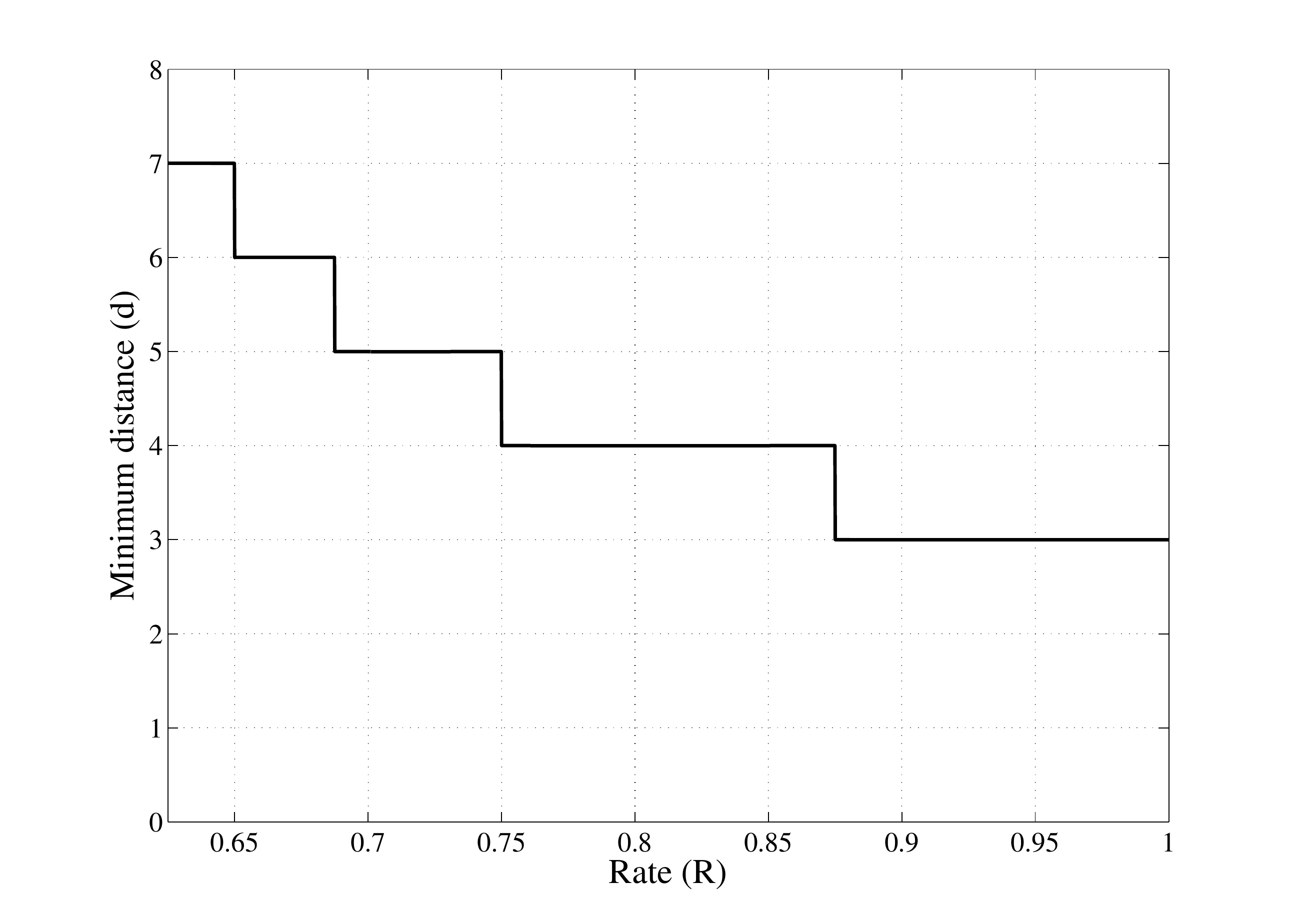}
  \caption{Minimum distance versus rate for planar codes.}
  \label{fig:dvr}
\end{figure}

A plot of the upper bound of Theorem \ref{thm:main} on $d$ versus $R$ for planar codes is shown in Fig. \ref{fig:dvr}.

The girth of $\mathcal{C}^\dagger (G)$ cannot be arbitrarily large. In fact, all $\mathcal{C}^\dagger (G)$ have girth $g \le 5$. This is because the girth of $\mathcal{C}^\dagger (G)$ is lesser than or equal to the minimum node degree in $\mathcal{C}(G)$, which is less than 6 by planarity; hence, $g$ can take a maximum value of 5. 

\begin{cor}
Let $G$ be a planar Tanner graph which supports a code of rate $R \ge \frac{9}{16}$ with the girth of $\mathcal{C}^\dagger(G)$ greater than or equal to 5.  Then,
\begin{align}
d \le  \left\lceil \frac{7-8R}{2(2R-1)} \right\rceil +3. 
\end{align}
\end{cor}
\begin{IEEEproof}
Its easy to see that the girth condition is met for all  $p \le 10$ (since $p=3.2^{l(p)-1} - 2 + z(p)$). With similar calculations as above, one can show that this places restriction on the rate as $R \ge \frac{9}{16}$.
\end{IEEEproof}

\section{Conclusion}
\label{sec:conclusion}
In this paper, we showed a bound on the minimum distance of high-rate ($\ge5/8$) codes that have planar Tanner graphs. The main result is the plot of the upper bound on minimum distance as a function of rate as shown in Fig. \ref{fig:dvr}. In particular, we see that such codes have a maximum minimum distance of 7. Hence, non-planarity is essential for the construction of codes on graphs with high minimum distance.

The proof uses ideas from graph theory, coding theory and an averaging argument through a series of constructions that exploit the planarity of a Tanner graph. Ideas from the proof could be possibly employed in construction of codes on non-planar graphs in the future.

Extending the bound to codes of all rates with a planar Tanner graph is an interesting problem for future study. We conjecture that codes with planar Tanner graphs will not support codes with large minimum distance for any rate.

\appendices
\section{Ordering nodes in $\mathcal{C}_U^\dagger(G)$}
\label{sec:numbering}
In this appendix, we show that $\mathcal{C}_U^\dagger(G)$ for a proper subset of nodes $U$ can be recursively constructed by adding the nodes from $U$ one at a time in an order such that each newly added node has degree 1 or 2. For an ordered set $U=\{u_1,u_2\cdots,u_{|U|}\}$, let $U_i=\{u_1,u_2,\cdots,u_i\}$.
\begin{prop}  Let  $\mathcal{C}_U^\dagger(G)$ be a connected subgraph of $\mathcal{C}^\dagger(G)$ induced by a proper node subset $U$. Then, there exists an ordering of the nodes in $U$, given by $U=\{u_1,u_2\cdots,u_{|U|}\}$, such that the degree of the node $u_i$ in $\mathcal{C}_{U_i}^\dagger(G)$ is either 1 or 2, and $\mathcal{C}_{U_i}^\dagger(G)$ is connected for $1\leq i\leq |U|$.
\end{prop}
To prove the proposition, we first claim the following:

\noindent\textit{Claim:} Let $\mathcal{C}_V^\dagger(G)$ be a subgraph of $\mathcal{C}^\dagger(G)$ induced by a proper subset $V$ of nodes. Then there exists either a degree-1 node in  $\mathcal{C}_V^\dagger(G)$ or a degree-2 node that is not a cut-vertex of $\mathcal{C}_V^\dagger(G)$.

\noindent\textit{Proof of claim:} We see that  $\mathcal{C}_V^\dagger(G)$ has at least one node $v$ of degree $\le 2$, since $\mathcal{C}_V^\dagger(G)$ is a proper subgraph of $\mathcal{C}^\dagger(G)$. If $v$ is a degree-1 or a degree-2 non-cut vertex, we are done. Otherwise, if $v$ is a degree-2 cut-vertex node, let $V_1$ and $V_2$ be the vertex sets of the two components of $\mathcal{C}_V^\dagger(G) - v$.

If all the nodes in $V_1$ are of degree 3 in $\mathcal{C}_V^\dagger(G)$, then the edge joining $v$ and $V_1$ will be a cut edge in $\mathcal{C}^\dagger(G)$. A cut edge in $\mathcal{C}^\dagger(G)$ implies a loop in $\mathcal{C}(G)$ \cite{harary}, which is not possible by construction. By similar arguments for $V_2$, we see that there is a node in $V_1$ and a node in $V_2$ with degree $\le 2$ in $\mathcal{C}_V^\dagger(G)$. Let $v_1$ be a node in $V_1$ of degree $\le 2$ in  $\mathcal{C}_V^\dagger(G)$. If the node $v_1$ is of degree 1 or if it is a degree-2 non-cut vertex, we are done. Otherwise, proceed with $v_1$ in place of $v$ and $\mathcal{C}_{V_1}^\dagger(G)$ in place of $\mathcal{C}_V^\dagger(G)$. Since the vertex set is finite, we are guaranteed to find a degree-1 or a degree-2 non-cut vertex proceeding to smaller components.

We are now ready to prove the proposition. 
\begin{IEEEproof} In $\mathcal{C}_U^\dagger(G)$, let $u_{|U|}$ be a degree-1 or a degree-2 non-cut vertex. For $2\leq i\leq |U|-1$, let $u_i$ be a degree-1 or a degree-2 non-cut vertex in $\mathcal{C}_U^\dagger(G) - u_{|U|} - u_{|U|-1} \cdots - u_{i+1}$. Observe that $\mathcal{C}_U^\dagger(G) - u_{|U|} - u_{|U|-1} \cdots - u_{i+1}$ is connected. The required ordering is then  given by $\{u_1,u_2,\cdots,u_{|U|}\}$.
\end{IEEEproof}

\section{Computation of $\eta_2$}
\label{sec:eta_2}
\begin{prop} Let $G'$ be the check inverse of a planar Tanner graph $G$. Let $\eta_2$ be defined as in (\ref{eta}). Then,
\begin{eqnarray}
\eta_2 = |V'^b_2|(\frac {4} {3} p + \frac {2}{3})t(p) - \sum_{e_l^\dagger \in \mathfrak{B}_o(G')}  r(e_l^\dagger)
\end{eqnarray}
\label{prop:q_deg2}
\end{prop}
provided the girth condition is satisfied for $\mathcal{C}^\dagger(G)$.
\begin{IEEEproof}
A degree-two bit node $v_b$ is identified with an edge $e$ in $\mathcal{C}(G')$. Since $e$ is common to two faces, say $f_1$ and $f_2$, $v_b$ is counted whenever either or both the corresponding nodes  (with same notation $f_1$ and $f_2$ as in Section \ref{sec:dual} )  in $\mathcal{C}^\dagger(G')$ is in the vertex set of complete 3-graphs on $p$ vertices. Let $e^\dagger$ be the edge connecting $f_1$ and $f_2$ in $\mathcal{C}^\dagger(G')$ corresponding to $e$ i.e, the edge that is occupied by $v_b$. The situation is depicted in Fig. \ref{fig:eta2}.
\begin{figure}[ht]
  \centering
  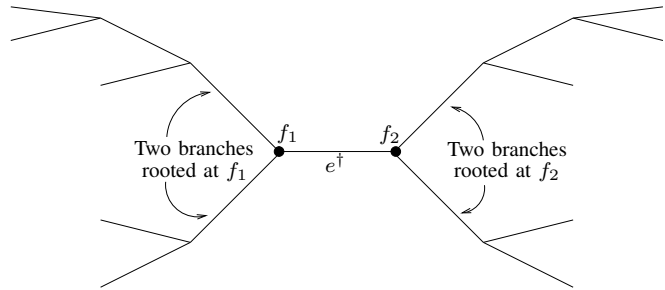
  \caption{Computing $\eta_2$.}
  \label{fig:eta2}
\end{figure}
Let $T_1$ and $T_2$ be the set of complete 3-graphs containing $f_1$ and $f_2$, respectively. Then,
$$q(v^b) = |T_1 \bigcup T_2| = |T_1| + |T_2| - |T_1 \bigcap T_2|.$$
Since $|T_1|=|T_2|=pt(p)$ (assuming the girth condition), we are left with computing $|T_1 \bigcap T_2|$. Note that $T_1 \bigcap T_2$ is the set of complete 3-graphs that contain both $f_1$ and $f_2$ i.e, containing $e^\dagger$ as shown in Fig. \ref{fig:eta2}. We will now count the number of 3-trees that will contain $e^{\dagger}$.
\begin{enumerate}
\item There are two branches rooted at $f_1$ and two at $f_2$ as seen in Fig. \ref{fig:eta2}. The number of nodes up to level $l(p)-2$ from the respective roots in these four branches is $4(2^{l(p)-2}-1)$. Every complete 3-tree rooted at any of these nodes will contain $e^{\dagger}$. 
\item The number of possible nodes in level $l(p)-1$ in the four branches is $4(2^{l(p)-2})$. Among the complete 3-trees rooted at these nodes, a fraction $\dfrac{z(p)}{3\cdot2^{l(p)-1}}$ will contain $e^{\dagger}$. 
\item Finally, all complete 3-trees rooted at $f_1$ and $f_2$ will contain $e^{\dagger}$. 
\end{enumerate}
Hence, the total number of complete 3-trees that contain $e^{\dagger}$ is given by 
$$\left(4(2^{l(p)-2}-1)+4(2^{l(p)-2})\dfrac{z(p)}{3\cdot2^{l(p)-1}}+2\right)t(p).$$
In addition, there will be $r(e^{\dagger})$ complete 3-graphs that contain $e^{\dagger}$ as a cycle-creating edge. Therefore,
\begin{align}
|T_1 \bigcap T_2| &= \left ( 2^{l(p)} + \frac{2}{3}z(p)  - 2 \right )t(p) + r(e^\dagger), \nonumber\\
                  &=(\frac{2}{3}p - \frac{2}{3})t(p) + r(e^\dagger). \label{intersection}
\end{align}
Hence, $q(v_b)$ for $v_b \in V_2^{'b}$ is
$$q(v_b) = (\frac {4}{3} p + \frac {2}{3})t(p) - r(e^\dagger).$$
Letting
\begin{equation}
\theta(p) = (\frac {4}{3} p + \frac {2}{3})t(p), \label{eq:theta} 
\end{equation}
we have
\begin{eqnarray*}
\eta_2&=&\sum_{v_b\in V_2^{'b}}\theta(p)-\sum_{e_l^\dagger\in\mathfrak{B}_o(G')}r(e^\dagger)\\
&=&|V'^b_2|\theta(p) - \sum_{e_l^\dagger \in \mathfrak{B}_o(G')}  r(e_l^\dagger),
\end{eqnarray*}

\end{IEEEproof}

\section{Computation of $\eta_1$}
\label{sec:eta_1}
\begin{prop} Let $G$ be a planar Tanner graph. Let $s_G$ be the number of singular nodes in $G$. Then, provided the girth condition on $\mathcal{C}^\dagger(G)$ is satisfied, 
\begin{equation*}
\eta_1  \ge |V'^b_1|(\frac {4} {3} p + \frac {2}{3})t(p) - s_G(p),
\end{equation*}
where $\eta_1$ is defined in (\ref{eta}) and 
$$s_G(p) = \begin{cases} 
  s_G, \;  p = 4,\\
  0, \;\text{else}.
\end{cases}$$
\label{prop:eta2}
\end{prop}
\begin{IEEEproof}
A degree-1 bit node $v_b$ in $G$ is connected to one check node, and is identified with that node in $\mathcal{C}(G)$. Every node in $\mathcal{C}(G)$ maps to a face in $\mathcal{C}^\dagger(G)$. Let $V^\dagger (v_b)$ be the set of vertices of $\mathcal{C}^\dagger(G)$ forming the face corresponding to $\mathcal{N}(v_b)$. Then $v_b$ contributes to the summation of weights whenever one or more of the vertices in $V^\dagger (v_b)$ is in the vertex set of complete 3-graphs. 

For $1\le p\le4$, the girth of $\mathcal{C}^\dagger(G)$ is at least 3. Two possible situations with girth 3 are shown in Fig. \ref{fig:eta1}.
\begin{figure}[ht]
  \centering
  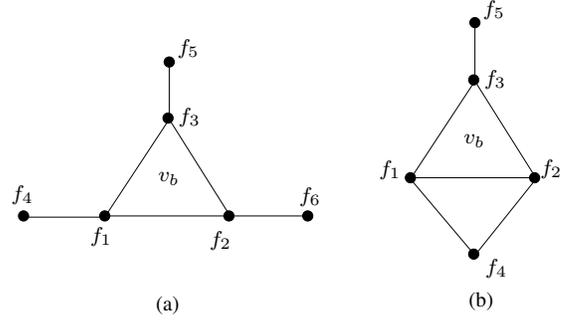
  \caption{Computing $q(v_b)$ for a degree-1 node $v_b$.}
  \label{fig:eta1}
\end{figure}
Table \ref{tab:qvb} enumerates $q(v_b)$ for $p=1,2,3,4$ for the two cases in Fig. \ref{fig:eta1} and compares with $\theta(p)=(\frac {4} {3} p + \frac {2}{3})t(p)$.
\begin{table}[ht]
  \centering
  \begin{tabular}{|c|c|c|c|}
    \hline
    $p$&Case from Fig. \ref{fig:eta1}&$q(v_b)$&$\theta(p)$\\
    \hline
    \hline
    1&(a),(b)&3&2\\
    \hline
    2&(a),(b)&12&10\\
    \hline
    3&(a)&15&14\\
    3&(b)&14&14\\
    \hline
    4&(a)&6&6\\
    4&(b)&5&6\\
    \hline
  \end{tabular}
  \caption{Enumeration of worst-case $q(v_b)$.}
  \label{tab:qvb}
\end{table}
From Table \ref{tab:qvb}, we see that the only case when $q(v_b)<\theta(p)$ is for $p=4$ in the situation of Case (b). Let $U=\{f_1,f_2,f_3,f_4\}$ in Fig. \ref{fig:eta1}. Then for the subgraph $\mathcal{C}^\dagger_U(G)$, the corresponding check graph $\mathcal{C}_U(G)$  is shown in Fig. \ref{fig:sing_check} as dashed lines. Hence, by the definition in Section \ref{sec:singular-nodes}, $v_b$ is singular. For $p\ge5$, we have girth at least 5 according to the girth condition. Hence, the face of $\mathcal{C}^{\dagger}(G)$ corresponding to $v_b$ has at least five nodes connected to other neighbours. The values of $q(v_b)$ in these cases can be readily seen to be much larger than $\theta(p)$ in a similar fashion.

\begin{figure}[htb]
\centering
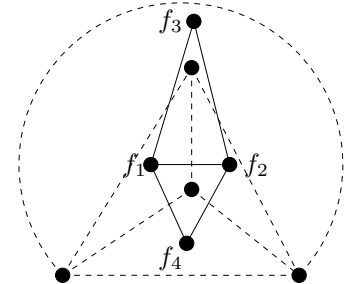
\caption{$\mathcal{C}_U(G)$ and $\mathcal{C}^\dagger_U(G)$ for singular case.}
\label{fig:sing_check}
\end{figure}

Therefore, we see that
\begin{equation*}
\eta_1  \ge |V'^b_1|(\frac {4} {3} p + \frac {2}{3})t(p)- s_G(p), 
\end{equation*}
where $$s_G(p) = \begin{cases} 
  s_G, \;p = 4,\\
  0, \;\text{else}.
\end{cases}$$  
\end{IEEEproof}

\section{Computation of $\alpha_x$}
\label{sec:gx}

\begin{prop}
Let $G$ be a planar Tanner graph with a fixed embedding. Let $\mathcal{Y}(G)$ be defined as in (\ref{eq:Ygen}). Let $\alpha_x$ be the increase in $\mathcal{Y}(G)$ when a new bit node of degree $x+3$ is embedded in $G$ such that the resulting graph is still planar. Then, 
\begin{equation*}
\alpha_1 \ge \left (\frac{2}{3}p - \frac{2}{3}\right )t(p),
\end{equation*}
and
\begin{equation*}
\alpha_2  \ge \left (\frac{1}{2}p - \frac{1}{6}z(p) - 1 \right)t(p).
\end{equation*}
where $p$ is such that girth condition is satisfied for $\mathcal{C}^{\dagger}(G)$.
\label{prop:gx}
\end{prop}
\begin{IEEEproof}
When $x=1$, a degree-3 bit node $v_b$ becomes a degree-4 bit node and $v_b$  is identified with two faces of the check graph having a common edge. This is equivalent to saying that $v_b$ is identified with an edge in the dual. So, $\alpha_1$ is equal to the number of complete 3-graphs that contain the edge. This number is already derived in (\ref{intersection}), and we get
$$\alpha_1 \ge \left (\frac{2}{3}p - \frac{2}{3}\right )t(p).$$
When $x=2$, a degree-3 bit node $v_b$ becomes a degree-5 bit node as shown in Fig. \ref{fig:g2}. The node $v_b$ is identified with three faces $f_1$, $f_2$ and $f_3$ of the check graph $\mathcal{C}(G)$. This corresponds to the connected subgraph $\mathcal{C}_U^\dagger(G)$ with $U=\{f_1,f_2,f_3\}$ in the dual as shown in Fig. \ref{fig:g2}.
\begin{figure}[htb]
\centering
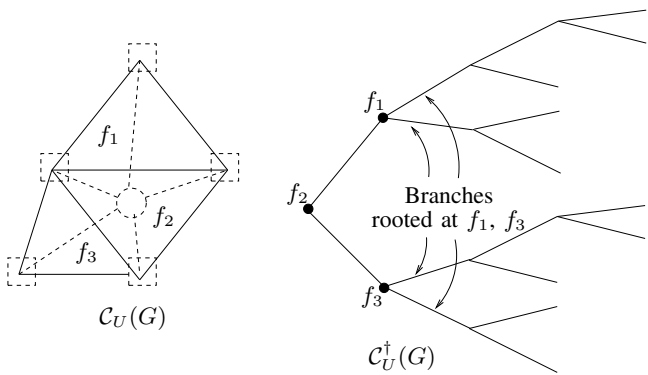
\caption{Computation of $\alpha_2$.}
\label{fig:g2}
\end{figure}
So, $\alpha_2$ is equal to the number of complete 3-graphs which contain $\mathcal{C}_U^\dagger(G)$. Let us now compute $\alpha_2$. All the complete 3-graphs rooted at nodes at a depth of $l(p) -2$ or lesser from the node $f_2$ in the branches shown in Fig. \ref{fig:g2} contain $\mathcal{C}_U^\dagger(G)$. Since girth condition is satisfied, there are $(3.2^{l(p)-2} -2)t(p)$ such complete 3-graphs. Now, consider only the child nodes of $f_1$ and $f_3$ in the branches shown in Fig. \ref{fig:g2} that are at a depth of $l(p)-1$ from $f_2$. Since girth condition is satisfied, the number of such nodes is  $2\cdot2^{l(p)-2}$ and there are $\frac{z(p)}{3\cdot2^{l(p)-1}}t(p)$  realizations of complete 3-graphs rooted at each such node containing $\mathcal{C}_U^\dagger(G)$. Thus,
 \begin{align*}
\alpha_2 & \ge \left( 3.2^{l(p)-2} - 2 + 2.2^{l(p)-2}  \frac{z(p)}{3.2^{l(p)-1}} \right )t(p), \\
 & \ge \left (\frac{1}{2}p - \frac{1}{6}z(p) - 1 \right)t(p).
\end{align*}
\end{IEEEproof}
\newpage
\bibliographystyle{IEEEtran}
\bibliography{docdb}

\begin{thebibliography}{1}
\providecommand{\url}[1]{#1}
\csname url@samestyle\endcsname
\providecommand{\newblock}{\relax}
\providecommand{\bibinfo}[2]{#2}
\providecommand{\BIBentrySTDinterwordspacing}{\spaceskip=0pt\relax}
\providecommand{\BIBentryALTinterwordstretchfactor}{4}
\providecommand{\BIBentryALTinterwordspacing}{\spaceskip=\fontdimen2\font plus
\BIBentryALTinterwordstretchfactor\fontdimen3\font minus
  \fontdimen4\font\relax}
\providecommand{\BIBforeignlanguage}[2]{{%
\expandafter\ifx\csname l@#1\endcsname\relax
\typeout{** WARNING: IEEEtran.bst: No hyphenation pattern has been}%
\typeout{** loaded for the language `#1'. Using the pattern for}%
\typeout{** the default language instead.}%
\else
\language=\csname l@#1\endcsname
\fi
#2}}
\providecommand{\BIBdecl}{\relax}
\BIBdecl

\bibitem{Gallager:1963xy}
R.~G. Gallager, \emph{Low-Density Parity-Check Codes}.\hskip 1em plus 0.5em
  minus 0.4em\relax Cambridge, MA: MIT Press, 1963.

\bibitem{Tanner:1981gd}
R.~M. Tanner, ``A recursive approach to low complexity codes,'' \emph{IEEE
  Trans. on Info. Theory}, vol.~27, no.~5, pp. 533--547, September 1981.

\bibitem{Etzion:1999jx}
T.~Etzion, A.~Trachtenberg, and A.~Vardy, ``Which codes have cycle-free tanner
  graphs?'' \emph{IEEE Trans. on Info. theory}, vol.~45, no.~6, pp. 2173--2181,
  Sep 1999.

\bibitem{Srimathy:2008qy}
S.~Srimathy and A.~Thangaraj, ``Codes that have tanner graphs with
  non-overlapping cycles,'' in \emph{5th International Symposium on Turbo Codes
  and Related Topics}, Sep 2008, pp. 299 -- 304.

\bibitem{Tanner:2001di}
R.~M. Tanner, ``Minimum-distance bounds by graph analysis,'' \emph{IEEE Trans.
  on Info. theory}, vol.~47, no.~2, pp. 808--821, Feb 2001.

\bibitem{Bondy:1976hl}
J.~A. Bondy and U.~S.~R. Murty, \emph{Graph Theory With Applications}.\hskip 1em
  plus 0.5em minus 0.4em\relax North-Holland, 1976.

\bibitem{harary}
F. Harary, \emph{Graph Theory}.\hskip 1em
  plus 0.5em minus 0.4em\relax Addison-Wesley Publishers, 1969.

\end{thebibliography}
\end{document}